\documentclass[12pt]{article}
\usepackage{amsmath}
\usepackage{times}
\usepackage{graphicx,subcaption}
\usepackage{color}
\usepackage{multirow}

\usepackage[authoryear]{natbib}
\usepackage{rotating}
\usepackage{bbm}
\usepackage{latexsym}

\newcommand{\captionfonts}{\normalsize}

\makeatletter  
\long\def\@makecaption#1#2{%
  \vskip\abovecaptionskip
  \sbox\@tempboxa{{\captionfonts #1: #2}}%
  \ifdim \wd\@tempboxa >\hsize
    {\captionfonts #1: #2\par}
  \else
    \hbox to\hsize{\hfil\box\@tempboxa\hfil}%
  \fi
  \vskip\belowcaptionskip}
\makeatother

\usepackage{cancel}
\usepackage{hyperref}
\newcommand{\mat}[1]{\ensuremath{\mathbf{#1}}}

\renewcommand{\vec}[1]{\ensuremath{\mathbf{#1}}}

\renewcommand{\vec}[1]{
\ensuremath{%
	\mathbf{#1}
}%
}

\newcommand{\Lk}{\ensuremath{\mathbf{L}^{\textrm{\scriptsize{-1}}}_{\textrm{k}}}}

\newcommand{\taustar}{\ensuremath{\overset{*}{\tau}}}

\newcommand{\ftilde}{\ensuremath{\tilde{f}}}

\newcommand{\C}{\ensuremath{\mathbf{C}}}
\newcommand{\M}{\ensuremath{\mathbf{M}}}

\newcommand{\ppre}{\ensuremath{\mathbf{p}^-}}
\newcommand{\ppost}{\ensuremath{\mathbf{p}^+}}

\newcommand{\ftildebold}{\ensuremath{\tilde{\mathbf{f}}}}

\newcommand{\pplus}{\ppost}
\newcommand{\pminus}{\ppre}

\newcommand\e[1]{%
  \ifcase#1\relax
    \textsc{x}%
  \or
    \textsc{y}%
  \or
    \textsc{z}%
  \or
    \textsc{q}%
  \or
    \textsc{r}%
  \or
    \textsc{s}%
  \or
    \textsc{u}%
  \or
    \textsc{v}%
  \or
    \textsc{w}%
  \fi
}
\newcommand\E[1]{%
  \ifcase#1\relax
    \mathrm{X}%
  \or
    \mathrm{Y}%
  \or
    \mathrm{Z}%
  \or
    \mathrm{Q}%
  \or
    \mathrm{R}%
  \or
    \mathrm{S}%
  \or
    \mathrm{U}%
  \or
    \mathrm{V}%
  \or
    \mathrm{W}%
  \fi
}

\renewcommand{\Lk}{\ensuremath{\mathbf{L}^{-1}_{k}}}

\DeclareMathAlphabet\mathbfcal{OMS}{cmsy}{b}{n}

\begin{document}
\hspace{13.9cm}1

\ \vspace{20mm}\\

{\LARGE Predicting the future with a scale-invariant temporal memory for the past}

\ \\
{\bf \large Wei Zhong Goh$^{\displaystyle 1}$, Varun Ursekar$^{\displaystyle 2}$, Marc W.~Howard$^{\displaystyle 2, \displaystyle 3}$}\\
{$^{\displaystyle 1}$Graduate Program in Neuroscience,}\\
{$^{\displaystyle 2}$Department of Physics,}\\
{$^{\displaystyle 3}$Department of Psychological and Brain Sciences,}\\
{Center for Systems Neuroscience,}\\
{610 Commonwealth Avenue,}\\
{Boston University.}\\


\thispagestyle{empty}
\markboth{}{NC instructions}
\ \vspace{-0mm}\\
\begin{center} {\bf Abstract} \end{center}
In recent years it has become clear that the brain maintains a 
temporal memory of recent events stretching far into the past. 
This paper presents a neurally-inspired algorithm to use a scale-invariant
temporal representation of the past to predict a scale-invariant future.  The
result is a scale-invariant estimate of future events as a function of the
time at which they are expected to occur.  The algorithm is time-local, with
credit assigned to the present event by observing how it affects the
prediction of the future.  To illustrate the potential utility of this
approach, we test the model on simultaneous renewal processes  with
different time scales.   The algorithm scales well on these problems despite
the fact that the number of states needed to describe them as a Markov process
grows exponentially.

\section{Using memory to predict the future}

Reinforcement learning (RL) models that are designed for Markov processes
\citep[e.g.,][]{WatkDaya92,Sutt88}
have been extraordinarily successful in accounting for 
reward systems in the brain \citep[e.g.,][]{SchuEtal97,WaelEtal01} and led to
remarkable achievements in artificial intelligence
\citep[e.g.,][]{MnihEtal15,SilvEtal18}.
Despite the success of RL, its affinity for Markov statistics may be a serious
limitation.  The real world contains many distinct causes that predict their
effects at a range of time scales, presenting a challenge for learners
optimized for Markov statistics.  Of course, random processes with memory can
be turned into Markov processes at the cost of defining additional states.
However, the cost in terms of memory, and time to learn transition
probabilities among an exponentially growing number of states, may be
excessively costly in some settings.  

It has been proposed that a primary function of the mammalian brain is to predict
future events to enable adaptive behavior \citep{Clar13,Fris10}.
Evidence from neuroscience has made
clear that the brain contains robust memory for the identity and time of
recent events extending well into the past.  
For instance, sequentially activated time cells in the hippocampus, prefrontal
cortex, and striatum \citep[e.g.,][]{MacDEtal11,TigaEtal18a,MellEtal15} maintain
information about the time at which recent events were experienced over at
least tens of seconds, and perhaps much longer.  
Experimental presentation of distinct stimuli triggers
different sequences of time cells \citep[e.g.,][]{TigaEtal18a,TaxiEtal20,CruzEtal20}
so that these populations
maintain information about what happened when.  In addition to
sequentially activated time cells, neurons in the entorhinal cortex
\citep{TsaoEtal18,BrigEtal20} and other cortical regions
\citep{BernEtal11,MurrEtal17} carry temporal information \emph{via} populations
of neurons that respond with a spectrum of characteristic time scales, in some cases up to at
least minutes \citep{TsaoEtal18}.   
This paper, inspired by work 
arguing that conditioning  results from an attempt to learn temporal contingencies between stimuli \citep{BalsGall09,GallEtal19},
presents a formal model that learns
to predict the future given a temporal record of the past.  This proposed mechanism
is computable given a temporal history that can be translated in time and
proposes a solution for how to estimate the future from a past that includes
information about many past events.

This paper proceeds as follows. In the rest of this section, we review a model for retaining a record of past events, and associations between event pairs. In Section~\ref{sec:creditassoc}, we present the model for predicting the future given a temporal record of the past. In Section~\ref{sec:properties algorithm}, we discuss its computational complexity, time scale invariance and several other properties. In Section~\ref{sec:Demonstration}, we present a numerical demonstration of the efficacy of this algorithm. Finally, in Section~\ref{sec:Discussion general}, we compare this algorithm to traditional RL algorithms, and point out its connections to neuroscience.

\subsection{A formal model for temporal record of the past}

We start with an agent which is capable of observing and remembering several
types of events, such as the onset of a 440~Hz tone or the appearance of an
image of an apple. In this
section, we will describe a model for its capabilities. We will see that the agent maintains a fuzzy timeline of past events, which it uses to make pairwise associations between events.
Neurobiological justification for this model is outlined in Section
\ref{subsec:appendix-formal-model} of the Appendix. 

\subsubsection{Events in continuous time}
\begin{figure}
    \centering
    \makebox[\textwidth][c]{\includegraphics[width=\textwidth]{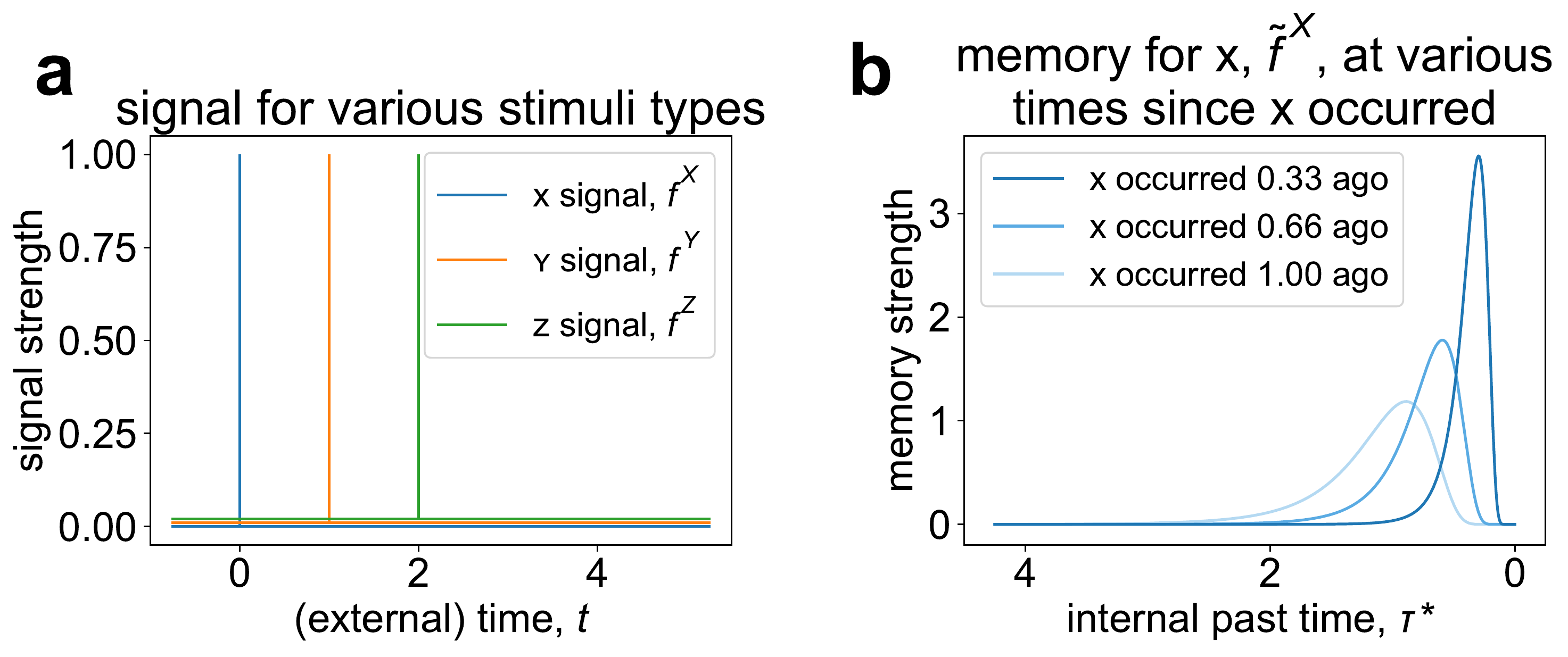}}
  \caption{\textbf{Memory is a fuzzy representation of the signal up to the
  present.} \textbf{(a)}~Signal as a function of external time, for three event
  types, \e{0}, \e{1} and \e{2}. This is the scenario considered in Fig.~\ref{fig:pairwise associations} through \ref{fig:credit density can differ}. \textbf{(b)}~Memory for a recent event as a function
  of internal past time, at varying (external) times since the event
  occurred. As a function of internal past time, peaks in the memory are
  present at approximately the time interval since the event.}
  \label{fig:signal and memory}
\end{figure}

We assume that the world provides a series of discrete events that occur in
continuous time. 
For simplicity, without loss of generality, suppose there are three types of
events, which we call \e{0}, \e{1} and \e{2} respectively. Whenever we need to avoid confusion, we will use \textit{event type} to refer to type of event, and use \textit{event episode} to refer to an individual occurrence of an event. We encode the occurrence of the event type \e{0} as a signal
$f^\E{0}(t)$, which is the sum of Dirac delta functions centered at the
occurrence times of episodes of \e{0} (Fig.~1a). (We will discuss quantities in relation
to \e{0}; such statements hold analogously for \e{1} and \e{2}.) We call $t$,
the argument for the signal $f^\E{0}(t)$, \emph{real time} or
\emph{external time}, to emphasize that this time axis is a feature of the world instead of being constructed by the observer. We denote the collection of all three
signals as $\vec{f}(t)$, and analogously for the quantities to follow. At
every instant in (external) time $t_0$, the agent has direct access to
$\vec{f}(t_0)$ (which is zero precisely unless the event of interest occurs at
$t_0$), but not $\vec{f}$ at any other time value.
Signals are shown in Fig.~\ref{fig:signal and memory}a in the case where
\e{0}, \e{1} and \e{2} occur at times 0, 1 and 2 respectively.  

\subsubsection{Temporal memory}
At every instant in time $t_0$, the agent's memory for \e{0}, denoted
$\ftilde^\E{0}(\taustar; t_0)$, is a fuzzy representation of the signal up to
the present, $f^\E{0}(t_0-\taustar)$. From the agent's perspective, the
internal past time, $\taustar>0$, indexes how long ago events in memory might
have occurred. The degree of fuzziness of the memory varies inversely with a
sharpness parameter $k$, which is typically a small even integer; throughout
this paper, it is fixed at~8. 

Consider an event that exactly happens at a particular time, $\tau_0$.  At time
$\tau_0 + t$, the memory element for that event is given by  
$\ftilde(\taustar;\tau_0+t) = \Phi_k(t/\taustar)/\taustar$, where the
fuzziness, $\Phi_k(\cdot)$, is given by the dimensionless equation
\begin{equation}
    \Phi_k(x)  = u(x)  \kappa_0 x^{k} e^{-k x} ,
    \label{eq:impresp}
\end{equation}
$\kappa_0=k^{k+1}/k!$ is a normalizing constant and $u$ is the unit step
function. Memories for a recent event are shown in Fig.~\ref{fig:signal and
memory}b for various values of $t$. For an arbitrary signal $\mathbf{f}$, the
associated memory up to time $t$ is 
\begin{equation}
	\ftildebold(\taustar; t) = \frac{1}{\taustar}\int_{-\infty}^t
	\mathbf{f}(\tau) \ \Phi_k\left(\frac{t-\tau}{\taustar}\right)\,d\tau .
    \label{eq:ftildeintegral}
\end{equation}
(The origin for Eqs.~\ref{eq:impresp} and \ref{eq:ftildeintegral} is outlined in Sec.~\ref{subsec:appendix-formal-model} in the Appendix.) In other words, the memory for an event type is the sum of the memory elements
associated with each episode of that event type.  
On its face, Eq.~\ref{eq:ftildeintegral} appears to assume that the agent has
access to the infinite past of $\mathbf{f}(t)$.  However, previous work has shown that
$\ftildebold(\taustar;t)$ can be efficiently and time-locally constructed from a
set of leaky integrators with a spectrum of time constants 
\citep[see~Section~\ref{subsec:appendix-formal-model} in the
Appendix;][]{ShanHowa13}.  
Using this approach, the number of leaky integrators necessary to remember the
past to some bound $T$ goes up like $\log T$.  Previous papers
\citep[e.g.,][]{ShanHowa13} using this formalism have made explicit use of this
property in choosing to sample values of $\taustar$ on a logarithmic scale and logarithmically compress the {\taustar} axis in integrals.
In this paper, we do not logarithmically compress the {\taustar} axis in integrals. However, one may adopt an alternative interpretation, consistent with this paper, as follows. Within this alternative interpretation, the expressions for integrals over {\taustar} used below are logarithmically compressed (i.e., a factor of $1/\taustar$ is added to the integrands). At the same time, the prefactor of $1/\taustar$ is removed from Eq.~\ref{eq:ftildeintegral}.  Neurally, this would amount to a
statement that the peak firing rate of time cells triggered by  a delta
function is constant as a function of $\taustar$.

The signal $\vec{f}$ up to any given external time $t_0$ fixes the event
occurrence history. However, due to the agent's fuzzy memory,
the agent is only able to form a fuzzy subjective belief distribution about
the event occurrence history leading up to the present. We may interpret the
memory for \e{0} as the agent's subjective estimate of the instantaneous rate
of occurrence of \e{0} at time $t-\taustar$. In other words, we have, for an
infinitesimal time element $d\taustar$, 
\begin{equation}
    \ftilde^\E{0}(\taustar;t)\,d\taustar \approx P\left( \e{0}\ @\ t-\taustar\ (d\taustar)\right),
\end{equation}
where $P(\cdot)$, the probability of an event, is used in the subjective
Bayesian sense to describe the agent's belief, and ``$\e{0}\ @\ t-\taustar\
(d\taustar)$'' stands for ``an episode of event \e{0} occurred within the infinitesimal
time interval between $t-\taustar$ and $t-\taustar+d\taustar$.'' 
Since $\ftildebold$ allows the agent access to the identity of and approximate time at which past events might have happened, we describe $\ftildebold(\taustar)$ to be a timeline of the past.

At each instant in time $t$, the agent is also able to compute the state of
the memory a time interval $\delta$ into the future, assuming that no events
of interest occur during that interval. For an arbitrary signal $\mathbf{f}$,
this quantity is given by 
\begin{equation}
    \ftildebold_\delta(\taustar; t) =
	\frac{1}{\taustar}\int_{-\infty}^t \mathbf{f}(\tau)\ \Phi_k\left(\frac{t+\delta-\tau}{\taustar}\right)\,d\tau .
    \label{eq:ftildedeltaintegral}
\end{equation}
Translation can be efficiently implemented based on the set of leaky integrators. Prior work has shown that this can be done in a
neurobiologically reasonable way
\citep[see Sec.~\ref{subsec:Appendix translation} in the Appendix;][]{ShanEtal16}.

\subsubsection{Estimating pairwise time-lagged statistics}

Many models of memory make use of associations between the temporal
context describing the recent past and the currently available stimulus. 
The agent described here builds pairwise associations from \e{0} (the cue) to \e{1} (the outcome) as the average state
of memory for \e{0} whenever \e{1} occurs, and analogously for other event
pairs: 
\begin{equation}
	\Delta M_\E{0}^{\E{1}}(\taustar) \propto 
    \ftilde^{\E{0}} (\taustar;t) f^\E{1} (t).
    \label{eq:M}
\end{equation}

We denote the collection of pairwise associations between event pairs as
$\M(\taustar)$, which may be thought of as an $n\times n$ matrix at every
$\taustar$, where $n$ is the number of possible events. We denote the
collection of pairwise associations with \e{0} as the cue as  $\M_\E{0}(\taustar)$, which may be thought of as a vector with $n$ elements, one for each possible outcome, at every $\taustar$. 

Note that as a neural network, Eq.~\ref{eq:M} simply requires Hebbian learning.
At the end of learning, we normalize $\M_\E{0}$ by the number of episodes of the cue
$\e{0}$, $\int f_\E{0}(t)\,dt$.

\begin{figure}    
\centering
\includegraphics[width=0.5\linewidth]{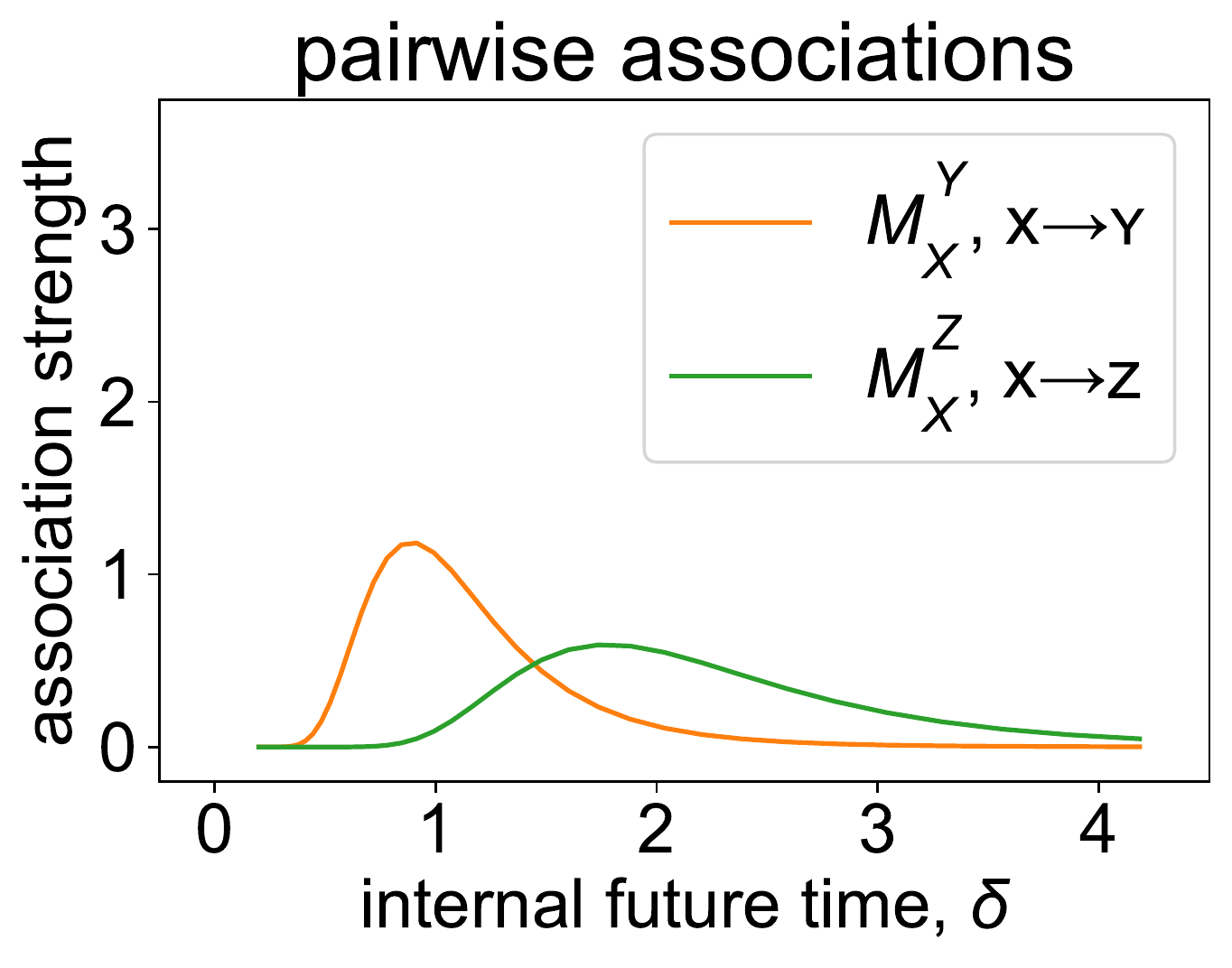}
\caption{\textbf{Pairwise associations fuzzily represent the rate of finding two
events occurring a certain time interval apart.} Other event types are ignored
in computing the association between two event types. The associations shown
here are based on the signals in Fig.~\ref{fig:signal and memory}a. As a
function of internal time, the associations peak at around the time interval
separating the event pairs.  }
\label{fig:pairwise associations}
\end{figure}

For example, suppose that \e{0} always precedes \e{1} by a time interval
$\tau_{\E{0}\E{1}}$. Then, by the end of learning, we would have the pairwise
association 
\begin{equation}
	M_\E{0}^\E{1}(\taustar) =  \Phi_k(\tau_{\E{0}\E{1}}/\taustar)/\taustar
    \label{eq:M in XY case}
\end{equation}
Fig.~\ref{fig:pairwise associations} shows the pairwise associations between two
pairs of events, occurring 1 and 2 time units apart respectively.

We may view $M_\E{0}^\E{1}$ from two complementary perspectives. Firstly,
given the occurrence of \e{1} in the present,
the agent
could use $M_\E{0}^\E{1}(\taustar)$
as a subjective estimate, based on an average over occurrences of $\e{1}$, of the instantaneous rate of occurrence of \e{0} at
time \taustar{} in the past, that is,
\begin{equation}
    M_\E{0}^\E{1}(\taustar)\,d\taustar \approx P\left( \e{0}\ @\  t_\E{1}-\taustar\  (d\taustar)\right).
    \label{eq: M interpret past}
\end{equation}
Secondly, given the occurrence of \e{0} in the present, 
the agent may use $M_\E{0}^\E{1}(\delta)$ as a subjective estimate
of the instantaneous rate of occurrence of \e{1} at time $\delta$ in the
future, that is,
\begin{equation}
    M_\E{0}^\E{1}(\delta)\,d\delta \approx P\left(\e{1}\ @\  t_\E{0}+\delta\ (d\delta)\right).
    \label{eq:M interpret future}
\end{equation}
We use $\taustar$ or  $\delta$ as the time argument for $\M$ according to
the interpretation that applies.

A limitation of directly selecting an element of the pairwise association $\M$ to predict the future is that the prediction can only be based on a single cue in the present (i.e., the cue corresponding to that element). To overcome this, the agent constructs the pairwise prediction $\mathbf{m}$, which integrates cue–outcome pairwise information (encoded by $\M$) from multiple simultaneous cues in the present to predict the future. We estimate
the rate of future occurrences of \e{1} based on the events in the present (time $t$) as
\begin{equation}
     m^\E{1}(\delta; t)\,d\delta \approx P\left(\e{1}\ @\ t+\delta\  (d\delta)\right),
     \label{eq:m as prediction}
\end{equation}
where
\begin{equation}
	\mathbf{m}(\delta; t) = \kappa_1e^{[\mathbfcal{M} \mathbf{f}_\delta](t)}
    \label{eq:m in terms of M operator}
\end{equation}
is the collection of rates of outcomes (a vector with $n$ elements, one for each possible outcome, at every $\delta$), and $\kappa_1$ is a normalization constant whose form is given in Sec.~\ref{subsec:Pairwise association and pairwise prediction}.
The exponential applies element-wise.
The operator $\mathbfcal{M}$ is defined by
\begin{equation}
    [\mathbfcal{M}\mathbf{f}_\delta](t) =
	\frac{1}{\left|\mathcal{E}_{t}\right|}\sum_{\alpha\in\mathcal{E}_{t}}\int
	f^\alpha_\delta (\taustar; t) \log \M_\alpha(\taustar)\,d\taustar,
    \label{eq:M operator definition}
\end{equation}
where $\mathcal{E}_{t}$ is the set of events occurring at time $t$,
$\left|\mathcal{E}_{t}\right|$ is the number of events occurring at time $t$,
$\M_\alpha$ is the collection of pairwise associations with $\alpha$ as the cue (a vector with $n$ elements, one for each possible outcome in correspondence with $\mathbf{m}$, at every $\delta$), and
\begin{equation}
    f_\delta^\alpha(\taustar; t) = \Phi_k(\delta/\taustar)/\taustar
    \label{eq:fdelta}
\end{equation}
denotes the future state of the memory element associated with the
currently occurring episode of $\alpha$.
(Contrast this with $\ftilde^\alpha_\delta$ (see Eq.~\ref{eq:ftildedeltaintegral}), which denotes the future memory state induced by all past occurrences of $\alpha$.)
The logarithm in Eq.~\ref{eq:M operator definition} applies element-wise.
The operator
$\mathbfcal{M}$ may be thought of as operating on the pre-computed future memory state of the
current events (Eq.~\ref{eq:ftildedeltaintegral}) to generate a prediction for the future. In general,
Eq.~\ref{eq:M interpret future} and Eq.~\ref{eq:m as prediction} provide
similar estimates for the future. The normalization constant $\kappa_1$ is such that
precisely when \e{0} is the only cue for \e{1} and the time delay between them is fixed, Eq.~\ref{eq:M interpret future} and
Eq.~\ref{eq:m as prediction} provide exactly the same estimate for \e{1}. Mathematical details can be found in Sec.~\ref{subsec:Pairwise association and pairwise prediction}.

Intuitively, the integral that is the second term on the right hand side of
Eq.~\ref{eq:M operator definition} takes the product of
$\log \M_\E{0}$ with the future state of the memory element $f^\E{0}_\delta$.
Let us consider $\e{1}$ to be the outcome of interest, and consider
the case where the time interval between $\e{0}$ and $\e{1}$,
$\tau_{\E{0}\E{1}}$, is constant. In this case, the integral (and thus,
$m^\E{1}$) attains a maximum as a function of $\delta$ when $\delta$ coincides
with the peak of  $ \log M_\E{0}^\E{1}(\taustar)$, which is approximately
$\tau_{\E{0}\E{1}}$. Fittingly, this is behavior we expect of $m^\E{1}$. 

Both $M_\E{0}^\E{1}$ and the corresponding Eq.~\ref{eq:M operator definition} integral are smooth functions that peak around the delay interval between the two events. 
This may prompt the question of why the integral is used instead of $M_\E{0}^\E{1}$ in its place \citep[e.g.,][]{TigaEtal19a}. 
The strength of the present formulation is that Eqs.~\ref{eq:m in terms of M operator}--\ref{eq:fdelta} closely parallel Eqs.~\ref{eq:prediction}--\ref{eq:C op definition} in the next section, for which the integral is necessary.
Using equations of similar form is more neurobiologically realistic, 
because it suggests that analogous neural architecture supports the computation for both pairwise prediction $\mathbf{m}$ and prediction $\mathbf{p}$, to be introduced later.

\section{Predicting the future with a scale-invariant past  \label{sec:creditassoc}}

It would be straightforward to build a prediction for the future based on a
single event (e.g., the most recent event) using the pairwise associations $\M$.
The challenge is to build a prediction that is based on multiple events in the
recent past. One difficulty arises when associations overlap. For example, we
associate the sound of rain (\e{0}) with a chance of hearing thunder (\e{2}). We also
associate the sight of wet ground (\e{1}) with a chance of hearing thunder (\e{2}). Having
heard the sound of rain, the prediction for thunder should not be increased by
the sight of wet ground when we step outdoors. This example illustrates one of
the pitfalls of simply adding the predictions suggested by the pairwise
associations.

To address double-counting, in addition to pairwise associations, we construct credit associations between event pairs, which is the key for this algorithm to generating a timeline for the future. In Sec.~\ref{subsec:Prediction from credit}, we explain how the agent constructs a timeline of the future by integrating over a timeline of the past, weighted by credit associations between cues and outcomes. In Sec.~\ref{subsec:Credit from prediction}, we show how the agent learns credit associations between cues and outcomes based on comparing predictions prior to the cue with predictions due to the cue.

\subsection{Generating predictions from credit associations}
\label{subsec:Prediction from credit}

In addition to the pairwise associations $\M$, we build the credit
associations $\C$ between each pair of events (a cue and an outcome) as a function of internal time
$\delta$ since the cue. The credit associations $\C(\delta)$ may be thought of as an $n\times n$ matrix at every $\delta$. 
We denote the collection of credit associations with \e{0} as the cue as $\C_\E{0}(\delta)$, which may be thought of as a vector with $n$ elements, one for each possible outcome, at every $\delta$. 
We interpret $C_\E{0}^\E{1}(\delta)$ as
logarithm of the factor by which an agent adjusts its subjective estimate of
the instantaneous rate of occurrence of \e{1} at time $\delta$ in the future,
having just observed \e{0}. 
Denoting $\pminus(\delta)$ as the agent's prior estimate
(just before observing $\e{0}$), we have 
\begin{equation}
	\left[p^-(\delta)\right]^\E{1} \exp C_\E{0}^\E{1}(\delta)\,d\delta \approx P\left(\e{1}\ @\  t_\E{0}+\delta\ (d\delta)\right).
    \label{eq:exp C prob interpret}
\end{equation}

Eq.~\ref{eq:exp C prob interpret} relates to the observation of one cue at one time (the present). For cues in the past, the further in the past they occur, the more imminent outcomes should seem. For example, if \e{0} has credit for \e{1} peaking at
$\delta=5$ and \e{0} occurred three time units ago, \e{1} should be expected in
two time units. Accounting for multiple cues over the past, we find
that at time $t$, the agent's internal timeline for 
a time $\delta$ into the future, is 
\begin{equation}
	\mathbf{p}(\delta; t) = \mathbf{\Lambda} \odot e^{[\mathbfcal{C}\ftildebold_\delta](t) },
    \label{eq:prediction}
\end{equation}
where $\mathbf{p}$ stands for \emph{prediction} and is a vector over event
types, $\mathbf{\Lambda}$ consists of the long-term average of each event type, $\odot$ denotes the element-wise product, and the exponential applies element-wise.
The operator $\mathbfcal{C}$ is defined by  
\begin{equation}
    [\mathbfcal{C} \ftildebold_\delta](t) = \sum_E \int  \C_E(\taustar) \ftilde_\delta^E(\taustar; t) \,d\taustar,
    \label{eq:C op definition}
\end{equation}
where the index of summation $E$ indexes the possible cue types, and $\C_E(\taustar)$ represents the collection of credit associations with $E$ as the cue (a vector with $n$ elements, one for every possible outcome, at every $\taustar$). 
Intuitively, the integral sums products of $\C_E$ with the projected memory
$\ftilde_\delta^E$.
Let us consider $\e{1}$ to be the outcome of interest and $\e{0}$
the only cue, and consider the case where the time interval between $\e{0}$
and $\e{1}$, $\tau_{\E{0}\E{1}}$, is constant. 
In this case, the integral (and thus, $p^\E{1}$) attains a maximum as a function of $\delta$ at the value of $\delta$ at which the peaks of $C_\E{0}^\E{1}$ and $\ftilde_\delta^\E{0}$ coincide. The credit
$C_\E{0}^\E{1}$ peaks around  $\tau_{\E{0}\E{1}}$, the time delay between
$\e{0}$ and $\e{1}$. The projected memory $\ftilde_\delta^E$ peaks around the
time $\tau_{\E{0}} + \delta$, where $\tau_{\E{0}}$ is the time that has
elapsed since $\e{0}$. Therefore, the integral (and thus, $p^\E{1}$) peaks
around $\delta = \tau_{\E{0}\E{1}} - \tau_{\E{0}}$, which is the time
remaining to $\e{1}$, the outcome of interest. In other words, the agent's
expectation for $\e{1}$ would be the highest at a time when $\e{1}$ is, in
fact, due. Mathematical details can be found in Sec.~\ref{subsec:Credit
association and prediction}.

We interpret $p^\E{1}(\delta; t)$ as the agent's subjective estimate, made at time $t$, of the instantaneous rate of occurrence of $\e{1}$ at time $\delta$ in the future, that is,
\begin{equation}
    p^\E{1}(\delta;t)\,d\delta \approx P\left( \e{1}\ @\  t+\delta\ (d\delta)\right).
    \label{eq:p prob interpret}
\end{equation}
Unlike Eq.~\ref{eq:M interpret future} and Eq.~\ref{eq:exp C prob interpret}, this estimate takes into account all of the events that have occurred in the recent past. 
A schematic distinguishing the utility of the pairwise associations $\M$ and the credit associations $\C$ in making predictions is shown in Fig.~\ref{fig:curly lines}. 
Just as we consider $\ftildebold(\taustar)$ a timeline of the past, we consider $\mathbf{p}(\delta)$ a timeline of the future.
Note that $p^\E{1}(\delta=0;t)$ would correspond to the agent's internal model for, in the language of point process theory, the conditional intensity function of \e{1} \citep[see][]{rasmussen_lecture_2018}. 

\begin{figure}    
\centering
\makebox[\textwidth][c]{\includegraphics[width=0.8\textwidth]{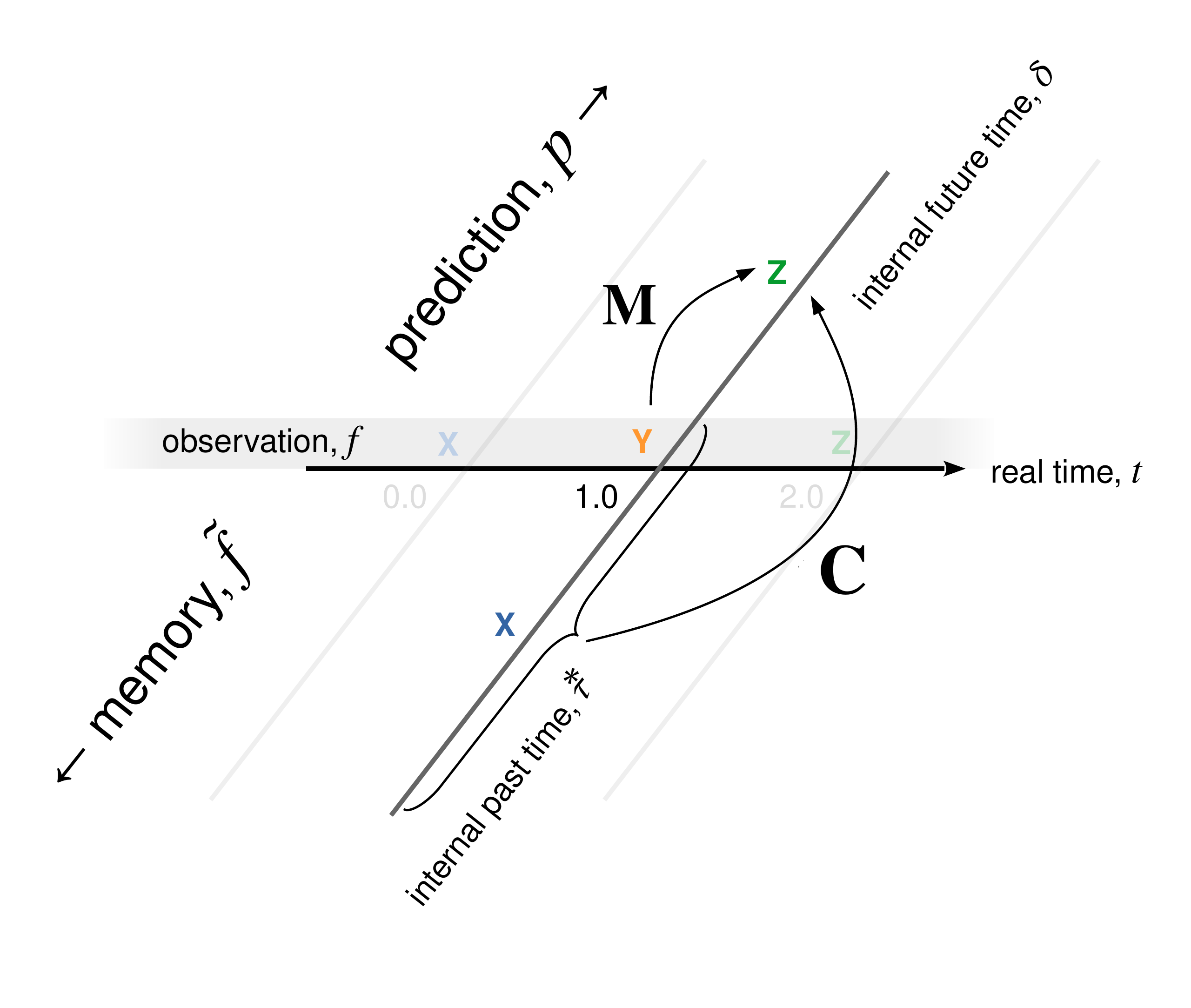}}
\caption{\textbf{Predictions can be made using credit associations $\C$ based
on memories of the multiple events in the recent past.} The horizontal axis
shows events occurring in real time. The event signal for this scenario is
shown in Fig.~\ref{fig:signal and memory}a. Associated with each point in real
time is an agent's internal time axis, shown here diagonally at $t=1.0$, which
indexes memories of the past (bottom half) and predictions for the future (top half).
The agent may make a prediction for the future with $\M$ (Eq.~\ref{eq:M
interpret future}) based on the currently observed event (here, \e{1}). As a
better alternative, the agent may make a prediction for the future with $\C$
(Eq.~\ref{eq:prediction}) based on multiple events in the present and the
recent past.  }
\label{fig:curly lines}
\end{figure}

As an illustration, consider again the scenario where events \e{0},
\e{1} and \e{2} always occur consecutively, once on each trial, at relative
times 0, 1 and 2 respectively, with a very long gap between trials. Once \e{0}
occurs, the proposed algorithm (explained in the following sections) generates
predictions for \e{1} and \e{2} that become more and more imminent as time
elapses (Fig.~\ref{fig:predictions become more imminent with time}). As a
function of $\delta$, the predictions peak at approximately the time when the
events are, in fact, due.

\begin{figure}    
\centering
\makebox[\textwidth][c]{\includegraphics[width=\textwidth]{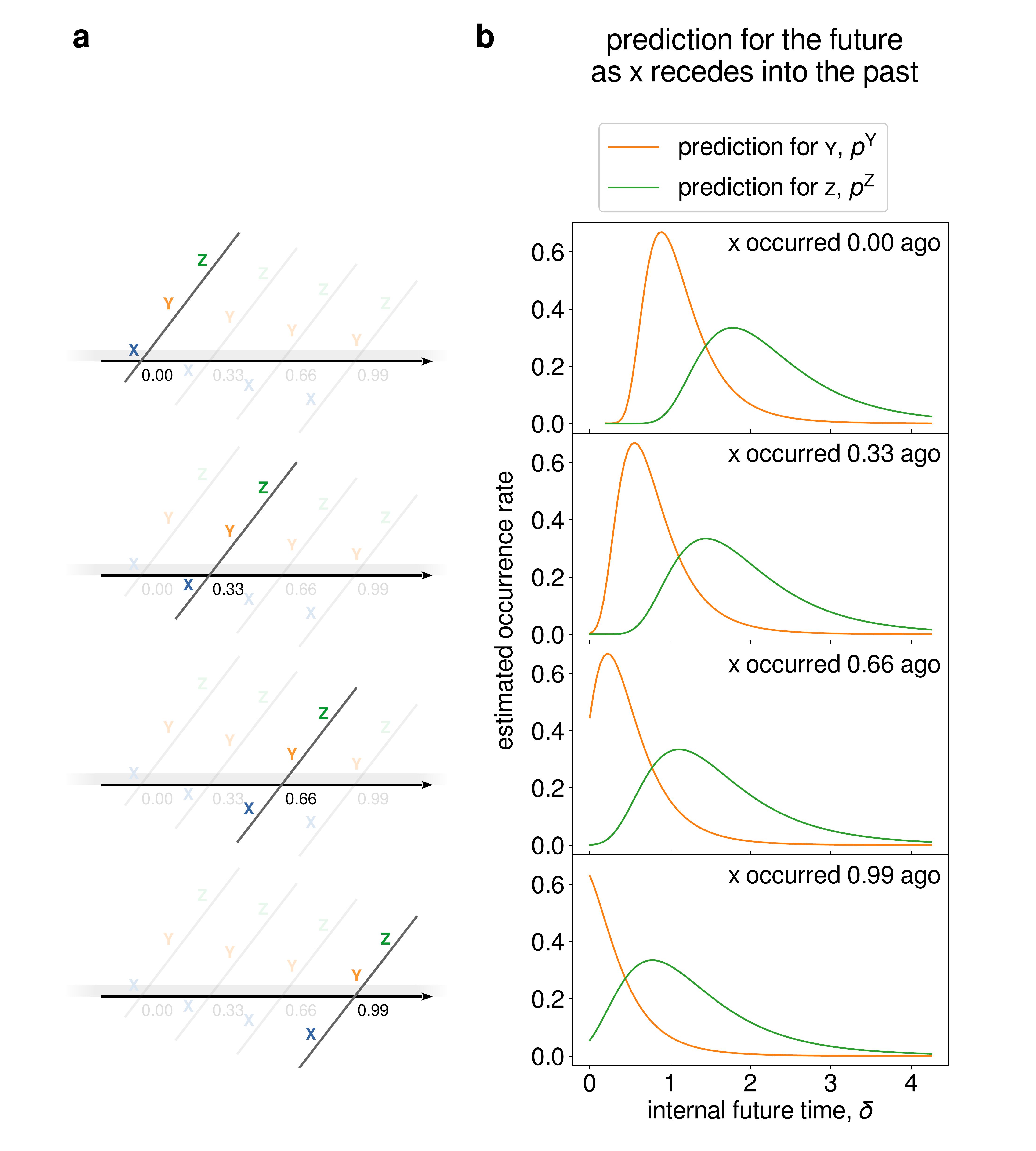}}
\caption{(Continued on the following page.)}
\label{fig:predictions become more imminent with time}
\end{figure}
\addtocounter{figure}{-1}
\begin{figure}
\caption{\textbf{Predictions for events peak at about the right time and
become more imminent with time.} The events \e{0}, \e{1} and \e{2} occur on
each trial at times 0, 1 and 2 respectively, as with previous figures. \textbf{(a)}~A schematic of the state of
memory and prediction as a function of time. The axes have the same
interpretation as in Fig.~\ref{fig:curly lines}. At real time 0.00, \e{0} is
observed, leading to a prediction for \e{1} and \e{2}, depicted along the
diagonal internal time axis. As real time passes, the memory of \e{0} recedes
into the past, and the prediction for \e{1} and \e{2} become more imminent,
depicted by the events' downward movement along the internal time axis. \textbf{(b)}~Prediction for \e{1} and \e{2} generated by simulation using the proposed
algorithm, as the memory for \e{0} recedes into the past, depicted at four
time points. The peak times for the prediction for \e{1} and \e{2} correspond
roughly to when the events are in fact due, and move towards zero as time passes. For
example, in the topmost plot, right after \e{0} occurs, \e{1} and \e{2} are to
occur in 1 and 2 time units respectively. Indeed, the generated predictions
for \e{1} and \e{2} peak at approximately $\delta=$ 1 and 2 respectively.  }
\end{figure}

\subsection{Computing credit associations}
\label{subsec:Credit from prediction}

Loosely speaking, we assign credit for an outcome to an event
according to how much the event's occurrence would revise the prediction for
that future outcome. In our example, wet ground would be assigned little to no
predictive value, since the chance of thunder has already been predicted by
the sound of rain.
During training, we update the credit assigned to an event when that
event occurs. In this section, we will describe the update that happens when
\e{0} occurs with no loss of generality.

Formally, as we have stated, $\exp C_\E{0}^\E{1}(\delta)$ is the factor by
which we should adjust the prediction for \e{1} at time $\delta$ in the
future, having just observed \e{0}. Therefore, to compute $\exp
C_\E{0}^\E{1}(\delta)$, whenever \e{0} is observed, we will first compute the
prediction for \e{1} before and due to the observation of \e{0}, and
analogously for other possible outcomes.

\subsubsection{Prediction prior to event observation}

Prior to event observation at time $t$, the prediction associated with internal future time $\delta$ is given simply by 
\begin{equation}
	\pminus(\delta; t) = \lim_{t'\rightarrow t^-} \mathbf{p}(\delta; t').
    \label{eq:pminus}
\end{equation}
This prediction arises from the memory of cues in the past, and specifically
excludes the effects of what occurs at time $t$.

Consider the scenario in Fig.~\ref{fig:signal and memory}, where \e{0}, \e{1}
and \e{2} occur consistently at trial times 0, 1 and
2 respectively. 
When \e{0} occurs, $(p^-)^\E{1}=\Lambda_\E{1}$, the long-term
average rate of \e{1}, for all $\delta$. This is because $\pminus$ is computed
based on memory of events occurring before \e{0}, of which there are none
(Fig.~\ref{fig:observed events receive less credit for predicted events}c). 
In contrast, when \e{1} occurs, $(p^-)^\E{2}$ shows a peak at $\delta=1$,
based
on memory of events occurring before \e{1} (i.e., \e{0}), and the credit
association between \e{0} and \e{2} 
(Fig.~\ref{fig:observed events receive less credit for predicted events}f).  

\subsubsection{Prediction due to event observation}
For the prediction due to the observed event \e{0} itself, we use the
pairwise prediction in accordance with 
Eq.~\ref{eq:m as prediction}, 
\begin{equation}
	\pplus(\delta; t) = \mathbf{m}(\delta; t).
    \label{eq:pplus}
\end{equation}
For the scenario in Fig.~\ref{fig:signal and memory},
when \e{0} occurs, $(p^+)^\E{1} = m^\E{1} =  M_\E{0}^\E{1}$
(Fig.~\ref{fig:observed events receive less credit for predicted events}c),
and when \e{1} occurs, $(p^+)^\E{2} = m^\E{2} = M_\E{1}^\E{2}$
(Fig.~\ref{fig:observed events receive less credit for predicted events}f).
Both of these have the same form, peaking sharply at $\delta=1$, since the
time interval between \e{0} and \e{1} and between \e{1} and \e{2} are fixed
and equal. 
 
\subsubsection{Updating \texorpdfstring{$\C$}{C}}
When \e{0} is observed at time $t$, we update $\C_\E{0}$ in the following manner:
\begin{equation}
	\Delta \exp \C_\E{0} (\delta)  \propto 
	\frac{\pplus(\delta; t)}{\pminus(\delta; t)} - \exp \C_\E{0} (\delta). 
    \label{eq:ppluspminus}
\end{equation}
The division and exponentiation are performed element-wise for each possible outcome.
This update depends on the previous state of $\C$ (through Eq.~\ref{eq:pminus} and Eq.~\ref{eq:prediction}). During training, as events occur, we
update respective components of $\C$, which in turn enhances the agent's
predictions of the future as training proceeds. This update rule squares with
the intuition that events be assigned credit in accordance with their
association with outcomes that are not previously predicted.
As training proceeds, $\exp \C_\E{0}(\delta)$ approaches $\pplus(\delta;t)/\pminus(\delta;t)$ in expectation, up to the variability of event occurrence history in recent episodes of \e{0} during training. 
Since we assume stationary statistics, a small learning rate (i.e., constant of proportionality in Eq.~\ref{eq:ppluspminus}) should be used to minimize the effects of such variability. 

For the scenario in Fig.~\ref{fig:signal and memory},
as noted, the observation of \e{0} generates a prior prediction for \e{2} that
is present when \e{1} occurs. Thus, via Eq.~\ref{eq:ppluspminus}, \e{1}
receives less credit for \e{2} than \e{0} for \e{1} at each $\delta$
(Fig.~\ref{fig:observed events receive less credit for predicted events}),
even though the \e{0}--\e{1} and \e{1}--\e{2} pairwise associations are the
same (Fig.~\ref{fig:credit density can differ}).
As a practical matter, since the learning of $\C$ depends on the accurate
learning of $\M$, for best results, $\C$ should be learned only after
$\M$ stabilizes during training.

\subsection{Summary}

The agent's memory $\ftildebold$ encodes a timeline of past events
(Eq.~\ref{eq:ftildeintegral}). Using Hebbian association, the agent makes
pairwise associations $\M$ between each pair of event types as a function of
internal time (Eq.~\ref{eq:M}). This lets the agent form a pairwise prediction $\mathbf{m}$ for
the future whenever events occur, but only based on the pairwise correlations
associated with those events as cues. To predict future events based on past events, the agent learns credit associations $\C$
between each pair of event types as a function of internal time. The agent
uses $\C$ and $\ftildebold$ to generate a timeline of future events
(Eq.~\ref{eq:prediction}). While the agent learns, each time an event occurs,
we step $\exp C^\beta_\alpha$ (where $\alpha$ is the event that occurred)
towards the ratio of the prediction for $\beta$ due to $\alpha$ (based on
$\M$), to the prediction for $\beta$ prior to $\alpha$ (based on $\C$)
(Eq.~\ref{eq:ppluspminus}). This design curbs double-counting of
correlations for an outcome associated with multiple cues at different points
in the past. Through learning, we expect the
agent to produce better and better predictions for events in its future.

\begin{figure}    
\centering
\makebox[\textwidth][c]{\includegraphics[width=1.2\textwidth]{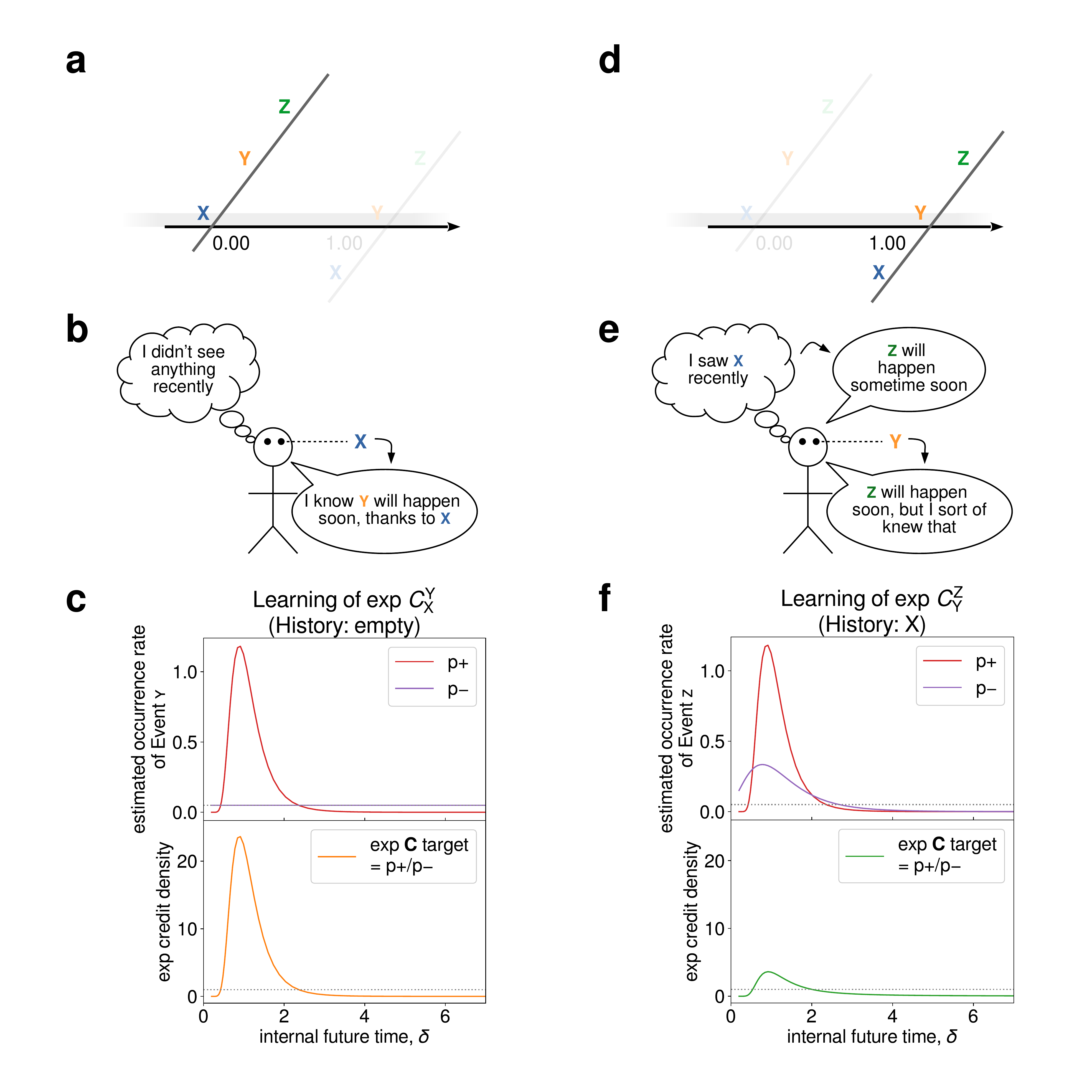}}
\caption{(Continued on the following page.)}
\label{fig:observed events receive less credit for predicted events}
\end{figure}
\addtocounter{figure}{-1}
\begin{figure}
\caption{\textbf{Observed events receive less credit for future events which
	have already been predicted based on past events.} As with all
	previous figures, the events \e{0}, \e{1} and \e{2} occur on each
	trial at times 0, 1 and 2 respectively. \textbf{(a)}~A schematic of
	the state of memory and prediction as a function of time, as in
	Fig.~\ref{fig:predictions become more imminent with time}(a). At real
	time 0.00, \e{0} is observed and the memory is empty. \textbf{(b)}~An
	illustration of an agent's inferences at the time \e{0} occurs. No
	memory of past events exists to suggest a prediction, whereas the
	currently observed event \e{0} suggests that \e{1} occurs soon.
	\textbf{(c)}~Plots of $\pplus$, $\pminus$ (top) and
	$e^{C_\E{0}^\E{1}}$ (bottom) as a function of internal future time,
	$\delta$, at the time \e{0} occurs, for the prediction of \e{1}. The
	quantity $\pplus$ (red) is the pairwise association between \e{0} and
	\e{1}, while $\pminus$ (purple) is flat as a function of $\delta$ as
	there is no memory of events. The quantity $e^{C_\E{0}^\E{1}} =
	\pplus/\pminus$ (orange). \textbf{(d)}~Same as (a), but at real time
	1.00. \e{1} is observed and \e{0} is in memory. \textbf{(e)}~An
	illustration of an agent's inferences at the time \e{1} occurs. The
	agent remembers \e{0}, prompting a prior prediction of \e{2}. The
	currently observed event \e{1} suggests the same, but the agent does
	not gain much information from \e{1}, and hence assigns \e{1} less
	credit. \textbf{(f)}~Same as (c), but at the time \e{1} occurs, for
	the prediction of \e{2}. The quantity $\pplus$ is the pairwise
	association between \e{1} and \e{2}, which is the same as that between
	\e{0} and \e{1}. However, $\pminus$ reflects the prior prediction for
	\e{2} based on the memory for \e{0}. (This is $p^\textrm{C}$ from the
	bottommost plot in Fig.~\ref{fig:predictions become more imminent with
	time}b.) Thus, $e^{C_\E{0}^\E{1}}$ is diminished.  }
\label{fig:construction of C}
\end{figure}

\begin{figure}    
\centering
\includegraphics[width=0.5\linewidth]{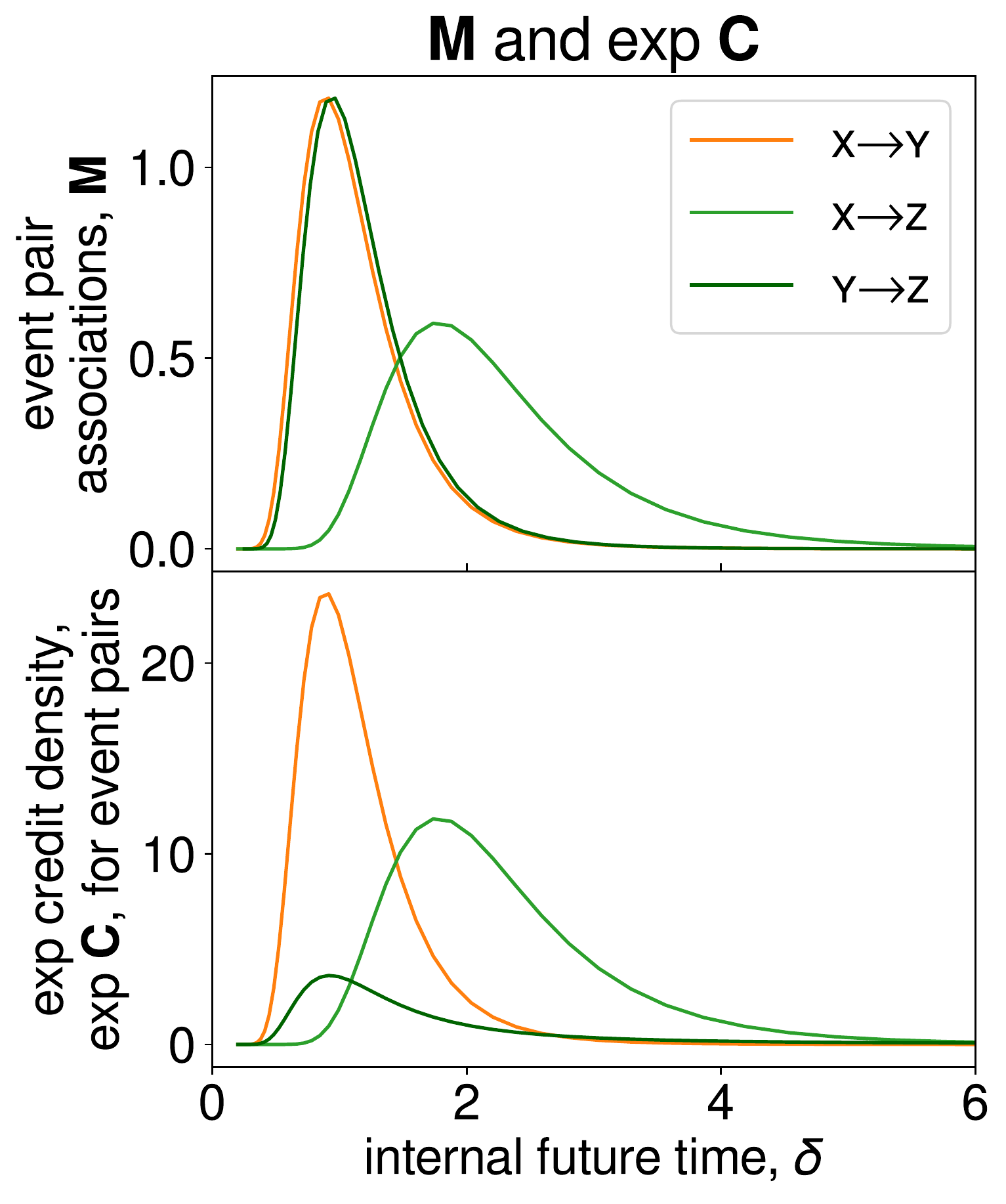}
\caption{\textbf{For event pairs, credit density can differ despite having the same pairwise associations.} A summary of the event pair associations (top) and credit densities (bottom) for all nontrivial event pairs for the scenario in all previous figures, where the events \e{0}, \e{1} and \e{2} occur on each trial at times 0, 1 and 2 respectively. The pairwise associations $M_\E{0}^\E{1}$ and $M_\E{1}^\E{2}$, overlapping perfectly, are slightly displaced for clarity. However, $C_\E{0}^\E{1}$ is greater than $C_\E{1}^\E{2}$ due to the memory of \e{0} allowing a prior prediction for \e{2} when \e{1} occurs, as shown in Fig.~\ref{fig:construction of C}(f).
}
\label{fig:credit density can differ}
\end{figure}

\section{Properties of the prediction algorithm}
\label{sec:properties algorithm}

The algorithm described above has interesting computational properties.  We
will discuss how it scales with the number of event types that can be distinguished
and the time scales over which prospection is implemented.  It can be shown
that the model is optimal for pairwise predictions modulo the 
uncertainty that comes from finite temporal resolution of memory.  Moreover,
the model is invariant to rescaling of time, which may be useful in
applications where the relevant time scale is not known
\emph{a priori}. 

\subsection{Scaling properties}
\label{subsec:Scaling properties computational complexity}

As with traditional associative models, the computational time and space
required for this algorithm vary quadratically with the number of event types
considered.  In typical RL models, each state $s$ must be
defined to include all of the information that could affect the transition to
the next state, in order to fit into the Markov structure. If transitions depend
on the indefinite past, then the number of possible
states would become unwieldy. In contrast, the event types used here are
economically defined to be those events that occupy a single point in time (\e{0},
\e{1}, etc.), which are much smaller in number.

In addition, this algorithm runs in time and space polynomial in the number
of $\taustar$ time points considered in $\ftildebold(\taustar)$ and $\delta$
time points considered in $\mathbf{p}(\delta)$. 
For example, in Eq.~\ref{eq:C op definition}, for each $\delta$, the numerical
integral is computed in time linear in the number of $\taustar$, the variable
of integration, corresponding to how far in the past memories are considered.
The full prediction, over all $\delta$ that the agent considers, is computed
in time linear in the number of $\delta$. Translation to different values of $\delta$ can either be implemented
serially, consistent with neural considerations \citep{ShanEtal16}, or be parallelized \emph{in silico}.
The quick performance comes at the cost of the ability to directly handle some
forms of joint statistics among cues. We discuss this
shortcoming in Sec.~\ref{subsec:limitations}. 

The longest time scale over which predictions are based and are
made increases exponentially with computational demands.   
Although the integral form in Eq.~\ref{eq:ftildeintegral} would seem to
require memory for the entire history up to the present, $\ftildebold$ can be
generated from leaky integrators with a number of time constants \citep{ShanHowa13}.
The scale invariance of $\Phi_k$ allows us to choose the distribution of time constants as a geometric series, resulting in a logarithmic relationship between the number of integrators and the longest time scale that can be represented.

\subsection{Equivalence of fuzzy memory and input temporal uncertainty}
\label{subsec:Equivalence of fuzzy memory and input temporal uncertainty}

Even when the time interval between events is fixed, fuzzy memory (finite $k$)
leads to temporal fuzziness in both the pairwise association $\M$ and
prediction $\mathbf{p}(\delta)$. At every instant in time, this induced fuzziness is
equivalent, in its effect on the prediction, to fuzziness due to intrinsic
temporal uncertainty in the signal $f$ faced by an agent with perfect memory
(infinite $k$).

As an example, consider an agent with fuzzy memory encountering $\e{0}$,
followed by $\e{1}$ after a fixed time interval $\tau$. Precisely at the time $\e{0}$ occurs, the agent's prediction for $\e{1}$ is given by
\begin{equation}
p^\E{1}(\delta; t_\E{0}) = \frac{\kappa_1^{-1}}{\delta}\Phi_k\left(\frac{\tau}{\delta}\right),
\label{eq:fuzzymemoryprediction}
\end{equation}
where $\kappa_1$ is as given in Sec.~\ref{subsec:Pairwise association and pairwise prediction}. Another agent
with perfect memory encountering $\e{0}$,  followed by $\e{1}$ after a random
time interval $\tau$, whose probability density function is given by
$q_\tau(t) = \Phi_k(t/\delta)/\delta$, makes an optimal prediction following
\e{0} equivalent to Eq.~\ref{eq:fuzzymemoryprediction}. The derivation of Eq.~\ref{eq:fuzzymemoryprediction} is given in Sec.~\ref{subsec:A.5.1 Worked example 1: Forward conditioning}.

Although the fuzzy memory agent's prediction for \e{1} some time after
encountering \e{0} is different from Eq.~\ref{eq:fuzzymemoryprediction}, this
equivalence property still holds: at every instant in time, there exists a
perfect memory agent, with observations subject to some density function of
$\tau$, with an equivalent optimal prediction.  

\subsection{Time scale invariance}

\begin{figure}    
\centering
\includegraphics[width=0.8\linewidth]{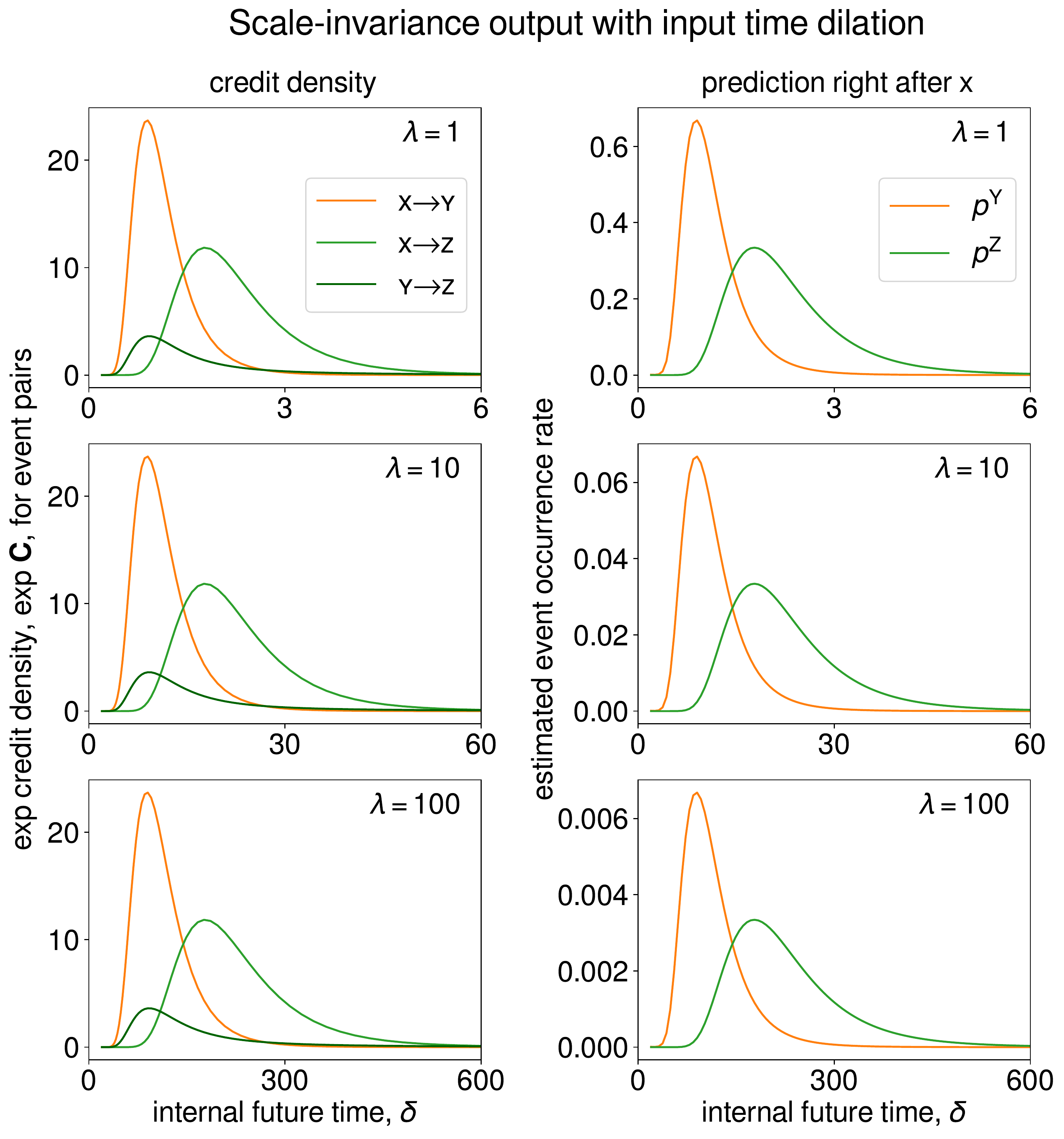}
\caption{\textbf{Predictions are time scale--invariant.} Top: Credit density and, as an example, the prediction after \e{0} occurs for \e{1} and \e{2}, as a function of internal future time, $\delta$, for the scenario in all previous figures. Middle, bottom: When the scenario is time-dilated, shown here by 10 and 100 times, the model output is unchanged as a function of dilated internal time. In the case of predictions, the magnitude rescales to preserve the area under the curve. This suggests that the proposed algorithm supports chains of events that occur over any time scale.
}
\label{fig:scale invariance}
\end{figure}

The prediction algorithm inherits the time scale invariance of the temporal
record of the past. If the input signals are time-dilated, the resulting
predictions would be time-dilated, rescaled in magnitude and otherwise
unchanged (Fig.~\ref{fig:scale invariance}). Therefore, the prediction
algorithm, with an appropriate range of $\taustar$ and $\delta$, supports
chains of events that happen over any time scale.  

Formally, for any constant $\lambda$,
the estimated probability of event occurrence within a small duration $d\delta$, 
$p\left(\delta; t\right) d\delta$, is invariant under the transformation
\begin{align*} 
t&\rightarrow\lambda t \\ 
\taustar&\rightarrow\lambda\taustar\\
\delta&\rightarrow\lambda\delta.
\end{align*}
This means that within the limits of a computational implementation, i.e., far
from the smallest and largest values of  $\taustar$  and $\delta$ (which grow
exponentially with the resources committed to representing time), the model
provides the same relative temporal resolution.

One may wonder whether, as an alternative to computing $\C$, one can generate
a future timeline $\mathbf{p}(\delta)$ and directly update it using $\pplus$
and $\pminus$ whenever an event occurs. A difficulty with this approach is
that a time scale would have to be chosen for the evolution of $\mathbf{p}(\delta)$
between events, violating the time scale invariance property that we desire. 

\subsection{With fuzzy memory, credit is assigned based on temporal proximity}
\label{subsec:With fuzzy memory, credit is assigned based on temporal proximity}

\begin{figure}    
\centering
\makebox[\textwidth][c]{\includegraphics[width=1.2\textwidth]{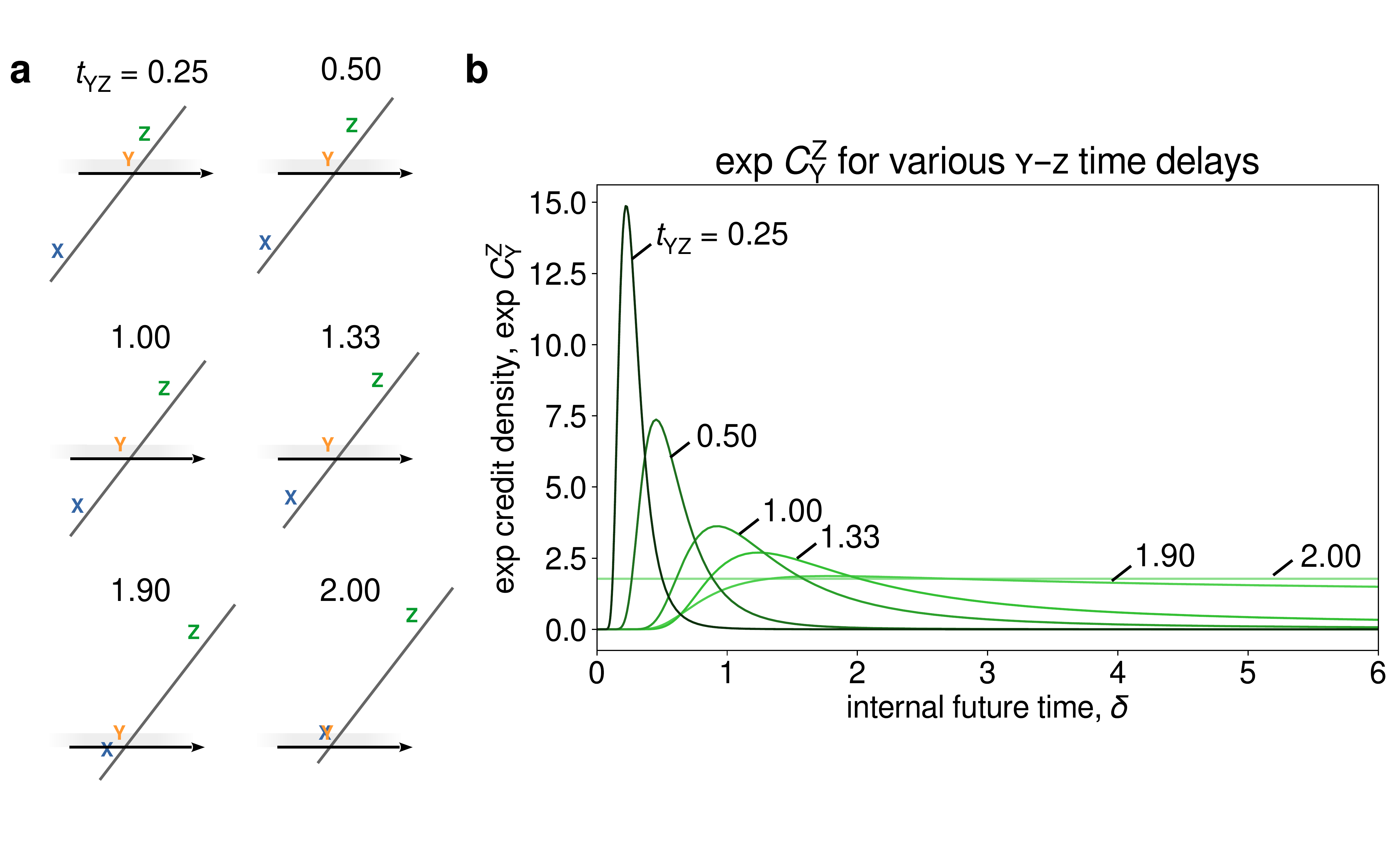}}
\caption{\textbf{Temporal proximity promotes credit assignment.} Events \e{0},
\e{1} and \e{2} occur at times 0, $2-t_{\E{1}\E{2}}$ and 2 respectively. \textbf{(a)}~A schematic of the state of
memory and prediction at the time \e{1} occurs, for six values of $t_{\E{1}\E{2}}$. The axes have the same
interpretation as in Fig.~\ref{fig:curly lines}.
\textbf{(b)}~Credit assigned to \e{1} for \e{2} is shown here for the six values of $t_{\E{1}\E{2}}$,
as a function of internal future time, $\delta$. In other words, each line
represents different amounts of temporal proximity between \e{1} and \e{2},
while the interval between \e{0} and \e{2} remains fixed. For
$t_{\E{1}\E{2}}=1.9$, \e{1} much closer in time to \e{0} than to \e{2}. In
this case, the credit is almost flat, as the prediction for \e{2} due to \e{0}
is still fresh. The case $t_{\E{1}\E{2}}=1$ is the scenario in
Fig.~\ref{fig:signal and memory} through \ref{fig:credit density can differ}. For lower and lower
values of $t_{\E{1}\E{2}}$, credit density is more and more sharply peaked.
The prediction for \e{2} due to \e{0} has flattened out, allowing the effect
of the pairwise association between \e{0} and \e{2} to dominate. The analytic form of the lines plotted are worked out in Sec.~\ref{subsec:A.5.2 Worked example 2: Credit and temporal proximity}.}
\label{fig:f5}
\end{figure}

Consider the scenario where $\e{0}$ occurs, then $\e{1}$, then $\e{2}$, always
with the same time delays.  In the limit of perfect memory, $\e{1}$ would
receive no credit for $\e{2}$. This is because the occurrence of $\e{0}$ would
allow the time of occurrence of $\e{2}$ to be predicted perfectly at all
times.  The occurrence of $\e{1}$ would not improve the (already perfect)
prediction. When memory is fuzzy, the \e{0}--\e{2} pairwise association would
have a larger temporal uncertainty than the \e{1}--\e{2} pairwise association,
since \e{1} and \e{2} are closer in time than \e{0} and \e{2}
(Eq.~\ref{eq:M}). Therefore, the occurrence of $\e{1}$ would improve the
prediction for $\e{2}$. The closer $\e{1}$ occurs to $\e{2}$, the more $\e{1}$
sharpens the prediction for $\e{2}$, and the more credit is assigned to
$\e{1}$ for $\e{2}$. Fig.~\ref{fig:f5} illustrates this effect, and supporting
equations are worked out in 
Sec.~\ref{subsec:A.5.2 Worked example 2: Credit and temporal proximity}.

\section{Demonstration: Event streams with memory and multiple characteristic time scales}
\label{sec:Demonstration}

We have seen that the algorithm described here is able to predict the
future based on a temporally extended record of the past containing multiple possible cues.  In addition, this prediction does not require selection of a preferred time scale, allowing for generalization across an exponentially large range of times.  As a consequence of these two properties, this approach is well-suited to applications where the relevant time scale is not known \emph{a priori} or to situations where there are multiple processes at different characteristic time scales that must be simultaneously learned.  To illustrate these properties, we demonstrate learning of the algorithm on a time series of discrete events generated from multiple Markov renewal processes (MRP).

In principle, the algorithm we describe is capable of handling multiple cues with additive effects (but see Sec.~\ref{subsec:limitations}) stretching into the indefinite past. 
However, for simplicity, we generate a scenario such that each event has exactly one cue. 
This cue is mostly found at most 15~time steps before the event. 
For comparison, most consecutive events have an intervening time of between 0.1 and 15 time units.
Crucially, the cue is not usually the immediately preceding event, but one of the several preceding events. 
Thus, one cannot merely predict the future based on the most recent event. 
To add realism, we introduce a small amount of variability in the event type of the outcome, 
as well as a small amount of Gaussian variability in the time of the outcome. 

\begin{figure}
    \centering
\includegraphics[width=1\linewidth]{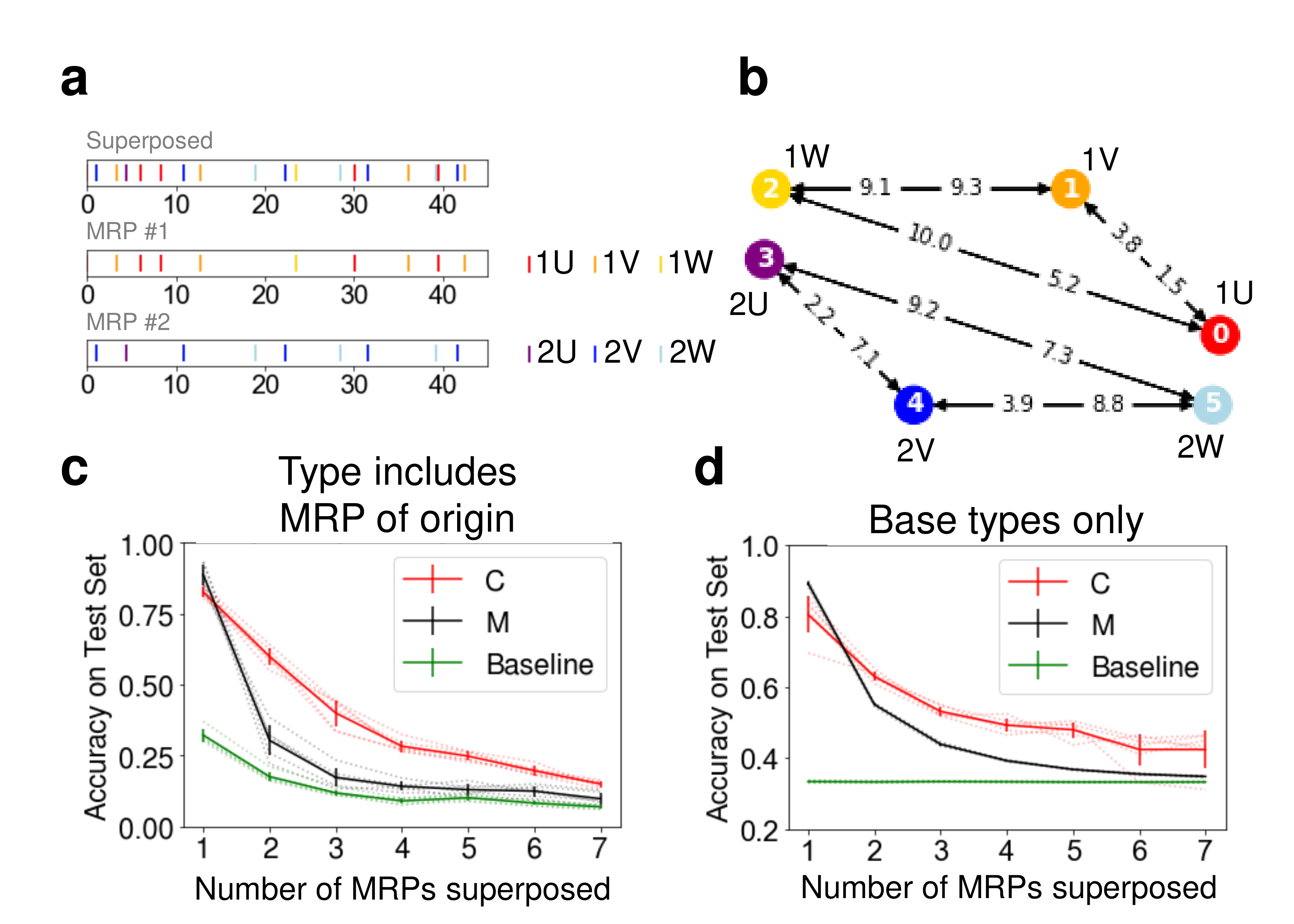}
  \caption{{\bf The algorithm provides good predictions for cues at multiple time scales.} \textbf{(a)}~The top panel shows the first few events in a superposed process. The bottom two panels show the corresponding events from the two component MRPs, which are composed of events $\{1\e{6},1\e{7},1\e{8}\}$ and $\{2\e{6},2\e{7},2\e{8}\}$, respectively.
  Note that successive events in the superposed process (e.g., the first two events in the topmost panel) may be from different MRPs, and thus the earlier event not predictive of the later event.
  \textbf{(b)}~Graph depicting mean transition times between event types within each component MRP. Weights are associated with the arrowhead closest to them. 
  The variances of the normally distributed transition times are not shown here.
  Note, for example, how the $1\e{7}\rightarrow1\e{6}$ transition takes place at the scale of about 1.5~time units, while the $1\e{6}\rightarrow1\e{8}$ transition takes place at a different scale of about 10 time units. 
  The two MRPs depicted here are two of the component MRPs in the simulation used to generate (c).
  \textbf{(c)}~We superpose MRPs such that event types from different MRPs are deemed different event types in the superposed process. \textbf{(d)}~We superpose MRPs such that event types from different MRPs are identified by base types (\e{6}, \e{7}, \e{8}) irrespective of their MRP of origin, resulting in exactly three event types in the superposed process.
  For (c) and (d), each point represents an average accuracy computed by repeating the training and testing procedures 6 times for each choice of number of MRPs superposed. Regardless of method of superposing MRPs,
  the algorithm (labeled $\mathcal{C}$) performs well above chance, showing that it provides good predictions for cues at multiple time scales. See text for a comparison of the $\mathcal{C}$- and $\mathcal{M}$-based predictions.
} 
 \label{fig:processesnwalkers}
\end{figure}

The way we generated a scenario with such properties is to superpose several MRPs, each with three base event types, $\e{6}$, $\e{7}$, $\e{8}$. 
MRPs have the property that the type of each event is the sole determiner of the probability distribution of the type and time of the next event. 
In other words, each event has a single cue. Superposing MRPs destroys the guarantee that the cue immediately precedes its outcome.
We generated the scenario using two approaches, mainly differing in the way event types are determined in the superposed process. 
For the first approach, event types in the superposed process are determined according to the base type of the event and the MRP of origin. 
For example, for a superposition of 7 MRPs, there would be $3\times7=21$ event types ($1\e{6},1\e{7},1\e{8},\dots,7\e{8}$). 
An example of such a scenario with two MRPs superposed is shown in Fig.~\ref{fig:processesnwalkers}a. 
Figure~\ref{fig:processesnwalkers}b shows the corresponding mean transition times for each type of transition. 
The drawback of this approach is that as the number of MRPs increases, the number of event types increases, making the prediction task inherently harder. 
For the second approach, event types in the superposed process are determined only according to the base types of the events, even if they originate from different MRPs. 
This way, for the prediction of the type of an event, there are always two wrong answers and one correct answer, for a fair comparison regardless of the number of MRPs superposed.

The algorithm we describe can be used to predict both the time and type of likely events in the future.
However, for simplicity, we evaluate the algorithm on its average accuracy of predicting the type of the next event, given the time to the next event, whenever an event occurs. 
We generate this prediction \textit{via} $\text{argmax}_i ~ p^i(\delta=t_{n+1}-t_n;t=t_n)$, where $t_n$ is the time of the $n$th event. We call this the $\mathcal{C}$-based prediction. 
As a comparison, we generate an $\mathcal{M}$-based prediction \textit{via} $\text{argmax}_i ~ m^i(\delta=t_{n+1}-t_n;t=t_n)$, where $j$ is the type of the event at $t_n$, and evaluate its average accuracy.
Notice that the $\mathcal{M}$-based prediction only invokes pairwise associations with event $j$ as the cue, whereas the $\mathcal{C}$-based prediction invokes credit associations with current and past events as cues.
Finally, we compare these to a baseline of always predicting the most frequent event type.
Our method is described in detail in Appendix~\ref{subsec:Methods for demonstration}.

The average accuracies of the prediction methods are shown in Fig.~\ref{fig:processesnwalkers}c and \ref{fig:processesnwalkers}d, as a function of the number of MRPs superposed, for the first and second approach of scenario generation respectively. 
The $\mathcal{C}$- and $\mathcal{M}$-based predictions generally outperform the baseline model.
Across both figures, the results are qualitatively similar. The accuracies of $\mathcal{C}$- and $\mathcal{M}$-based
predictions are comparable for a single MRP.
This is expected since for an MRP, the cue and its outcome are neighbors. Whenever an event occurs, the $\mathcal{M}$-based predictor uses the pairwise associations between that event and its possible outcomes to predict the type of the next event. 
However, as more and more MRPs are superposed, the $\mathcal{C}$-based algorithm outperforms. 

What drives the difference in performance between the $\mathcal{C}$- and $\mathcal{M}$-based algorithms? 
Although the $\mathcal{C}$-based algorithm uses the credit associations $\C$ while the $\mathcal{M}$-based algorithm uses the pairwise associations $\M$, this difference is immaterial in this case.
Since each event only has one cue, $\exp \C^\alpha$ is proportional to $\M^\alpha$.
Rather, the $\mathcal{M}$-based algorithm suffers when successive events originate from separate MRPs, and the pairwise association between the respective event types would not be predictive.
The $\mathcal{M}$-based algorithm makes predictions only based on events in the present. In contrast, the $\mathcal{C}$-based algorithm makes predictions based on events in the present and in the past, where the correct cue would be included in such situations. 

This demonstration provides a proof of concept that the algorithm provides reasonable predictions for cues at time scales spanning one order of magnitude. 
We accomplished this without selecting any single operating scale. 
The demonstration gives a flavor for the advantages of the algorithm we describe over Markov models.
A classic approach based on $n$-th order Markov models
would entail discretizing time at some lowest-level scale \citep[but
see][]{KurtRedi09,LudvEtal08}, and sizing the memory buffer to encompass most of the longest transitions. 
For simplicity, we have constructed a relatively tame scenario for this demonstration, in which most event relationships only span about 1--15~time units, and events are sparse.
In reality, the wider the range of time scales, the harder it is for standard algorithms operating at the lowest-level time scale, which
fumble at time scales significantly different from
their operating scale \citep{Moze92}. 
In scenarios where events have long-range
temporal dependencies, Markov models would be significantly limited
by the exponential growth in the number of states (and thus, computational demands) with the size of the memory
buffer. The algorithm we describe does not face these limitations (see Sec.~\ref{subsec:Scaling properties computational complexity}).

\section{Discussion}
\label{sec:Discussion general}

We have proposed an algorithm that generates a scale-invariant timeline of the
future. This algorithm is time-local in the sense that predictions at time $t$
are derived from $\ftildebold(\taustar, t)$, which represents events that are, in
fact, non-local in real time.   Moreover, the translation mechanism enables
event rates at future time points to be estimated.
In addition to
associative memory, as developed by model-free RL algorithms, this capability
would let an agent construct the estimate over possible futures
\citep{McGuKabl13}.  

\subsection{Theoretical properties of the current model}

This model has properties that are quite different from traditional
RL paradigms.
First, this algorithm naturally runs in continuous time, which suits
applications dealing with natural processes unfolding in time. This feature
contrasts with basic RL algorithms, which only allow
agents to move among discrete states in discrete time. In principle, this proposed
algorithm can be extended such that position in higher-dimensional spaces replaces or augments time, allowing
agents to navigate real and abstract spaces.
Translation can be along an angle or perhaps even along a trajectory, instead of being confined to a given axis (Eq.~\ref{eq:ftildedeltaintegral}).  

Second, the scale invariance of the model is useful in applications where the
time scale of event relationships is not known in advance. In principle, the model is
indifferent towards the absolute time intervals between events. Instead,
within a given scenario, it is only concerned about time intervals relative to
other time intervals. In comparison, in traditional RL
systems, a time scale for history dependency, if any, is set by the size of
the history that the designer defines as part of the state $s$. Moreover, in
many aspects of the world that we might be interested in, such as in natural
language \citep{AltmEtal12}, network traffic \citep{CoheEtal00} and financial
markets \citep{cont_long_2005}, event dependencies exist simultaneously across
a wide range of scales. This model is potentially suited for such applications,
since it incorporates past events across a range of time scales, and an
increase in computing resources provides an exponential increase in the length
of history considered.

Third, in the context of RL, this model
may be incorporated into algorithms to allow agents to naturally form a
prediction of its own trajectory as a function of future time. This can be done by considering the agent's arrival at
some or all states as events.  
In addition, by combining the predictions for future states $s$ as a function
of future time, $p^{s}(\delta)$, with a reward function over future states,
the agent can generate the predicted future reward as a function of future
time, $r(\delta)$. By learning and comparing weighted integrals of $r(\delta)$ for several alternative policies, the agent can choose flexibly among these policies according to task demands. For instance, if the agent knows it only has 10 time units to complete the task, it can choose the policy with the highest $\int_0^{10} r(\delta)\,d\delta$.  
The model's ability to form a
prediction as a function of future time stands in contrast to RL
paradigms, which tend to flatten the dimension of future time. For
example, a naive RL agent assigns values to states
according to the expected \emph{sum} of future reward starting from that
state; a successor representation agent \citep{Daya93} learns the expected
future state occupancy, \emph{summed over future time}, starting from each
state \citep[but see][]{TanoEtal20,MomeHowa18}.

Finally, we note that this model provides information usually associated with
model-based RL, but with very different computational properties.  Like
model-based RL, this model provides an explicit prediction as a function of
future time $\delta$.  However, a constraint of model-based RL is that the
time to compute an event $\delta$ in the future goes up linearly with
$\delta$.  In the present model, because the calculation of the prediction at
a particular value of $\delta$ does not depend on the prediction at previous
values of $\delta$, one could in principle compute all values of $\delta$ in
parallel.  Moreover, this means that it is possible to sample the  $\delta$
axis in whatever way is convenient.  Integrals over $\delta$ give hyperbolic
discounting if the $\delta$ axis is sampled evenly as a function of $\log
\delta$.  See also \citet{ShanEtal16} for considerations related to physically
instantiating translation across a population of neurons.

\subsection{Theoretical limitations of the current model}
\label{subsec:limitations}

We highlight two limitations relating to applying this algorithm toward
machine learning. First, the algorithm, as currently described, is not
directly sensitive to joint statistics of two or more cues. For example, the
model would be unable to capture the conditional structure ``\e{2} occurs
exactly if either \e{0} or \e{1} occurs, but not both''. As a consequence, the
algorithm is also unable to deal appropriately with number of events.  For
instance, the algorithm has no basis to differentiate ``\e{0} precedes \e{1}
by 10~s'' from  ``half of the time, \e{0} precedes two closely-spaced
occurrences of \e{1} by 10~s and the other half of the time, \e{1} does not
occur''.  We can mitigate this issue by perceiving events depending on
context.  For example, the agent can perceive the \e{1} after an \e{0} as the
event \e{0}\e{1}, enabling sensitivity to joint statistics of at most two
cues. In terms of computational complexity, naively implementing this would
introduce a quadratic factor in the number of base events. However, we can
reduce the resource complexity by finding a compressed representation of the
event history while preserving information about future events: that is,
dealing with the information bottleneck problem \citep{TishEtal00}.  Since
existing deep neural network algorithms efficiently extract joint statistics,
it would be natural to pursue research that seeks to merge this approach with
deep network algorithms.

Second, this algorithm is limited in prescribing how
to achieve optimal policies in the context of RL. Our
focus has been on how to predict future events, and not how to learn the best
policy. In many contexts, it is natural to define events such that events
occur depending on actions of the agent (e.g., in a spatial navigation task
where events occur based on the agent's trajectory). In these cases, in
effect, we presume that the agent follows an existing policy $\pi$, and the
model deduces event associations and makes predictions with respect to $\pi$.
The agent can certainly flexibly choose among several alternative policies,
say, between $\pi$ and $\pi'$, by comparing predictions from the start state
and selecting the more rewarding alternative. However, unlike basic
RL algorithms, we do not prescribe a method for learning a
policy that scales in complexity with the number of states, such as a policy
to navigate a grid. In the context of grid navigation, we have, in effect,
avoided assigning values to coordinates on the grid, since this contradicts
our design principle of allowing history to influence events (rewards). More
research would be needed if one wished to pursue policy learning within the
framework we describe.

\subsection{Neuroscience considerations}
This subsection discusses two potential points of contact between the formal model presented in this paper and computational and systems neuroscience.

\subsubsection{Reward prediction error and dopamine}
The success of RL algorithms in accounting for the firing of dopaminergic
neurons in the basal forebrain \citep{SchuEtal97} is arguably the greatest
achievement in computational cognitive neuroscience.  The basic empirical
story is well-known.  Dopaminergic neurons respond to unpredicted rewarding
outcomes.  However, with learning, as the reward becomes predicted by a
neutral stimulus, the cells no longer fire to the predicted reward but instead
fire to the neutral stimulus that predicts the future rewarding outcome
\citep[see][for a review of the early literature]{Schu06}.
While there are undoubtedly many details that would need to be worked out, at least the rough outline
of the classical empirical story about dopamine  can be mapped onto this framework.  

Let us suppose that expected future value is computed at each moment by
integrating over future time $\delta$, taking the projection from the vector of predicted events  $\mathbf{p}(\delta)$
onto some vector $\mathbf{A}$ that describes the reward value of each possible event type in  $\mathbf{p}$:
\begin{equation}
V(t) = \int_0^\infty \mathbf{A} \cdot \mathbf{p}(\delta; t)g(\delta)\,d\delta,
\end{equation}
where $g(\delta)$ denotes the factor arising from compression of the $\delta$ axis. As discussed above, it is reasonable to sample $\delta$ on a logarithmic scale
to implement hyperbolic discounting, in which case $g(\delta)=1/\delta$.  Let us suppose further that reward
prediction error $E$ is computed as the difference between $r(t)$, the actual reward
observed at time step $t$ and the change in $V$ at time step $t$:
\begin{equation}
	E(t) = r(t) +  \left[V(t) - V(t-\Delta t) \right],
\end{equation}
where we have chosen a discrete time interval $\Delta t$ to acknowledge that
the computation of value may take a substantial amount of time in the brain.
For instance, \citet{ShanEtal16} proposed that integrals over $\delta$ could
be completed within a theta oscillation,  suggesting $\Delta t$ could be as
long as a few hundred milliseconds. Now, consider slowly learning
an association between an inherently neutral event \e{0} and a rewarding event \e{1}, separated by a fixed delay $\tau$.
Initially, \e{1} is unpredicted.  When \e{0} is presented, there is no change
in $V$.  Similarly, $V$ is zero both before and after \e{1} is
presented.  Because \e{1} is rewarding, the reward prediction error is positive around the time of
presentation of \e{1}.  After learning, immediately after \e{0} is presented,
$\mathbf{p}(\delta)$ includes  the prediction for
\e{1}, a time $\delta \simeq \tau$ in the future.
This means that $V(t)$ changes abruptly around the time that \e{0} is
presented, resulting in a positive reward prediction error.  Now, after
learning, consider a time $\tau$ after presentation  of \e{0}.  If the
rewarding stimulus is omitted, negative reward prediction error is observed as
the peak in $\mathbf{p}(\delta)$ corresponding to \e{1} becomes increasingly
truncated.  However, if \e{1} is presented at the time it is expected, then
the positive reward from \e{1} is balanced by a rapid decreasing $V(t)$.  Note
that because \e{1} does not predict itself at a short lag, observation of
\e{1} abruptly decreases the prediction of itself.

This approach aligns well with the classic understanding of reward prediction
error with one very important exception.  Rather than estimating expected
future reward \emph{via} temporal difference learning, predictions for an
extended future are available at each moment.  Unlike temporal difference
learning algorithms, there is no sense in which value moves gradually along
intermediate time points between \e{0} and \e{1}.   This model thus has no
difficulty accounting for the finding that value seems to rapidly ``jump''
between events \citep{PanEtal05}.

\subsubsection{Translation and theta oscillations}
The algorithm described here relies on the ability to translate $\ftildebold$
towards the past.  \citet{ShanEtal16} suggested that hippocampal theta
(4--12~Hz) oscillations could provide a mechanism for translation of temporal
representations.  The basic conjecture of that model for translation is that
different values of $\delta$ map onto different phases of theta oscillations.
If the timeline $\delta$ maps onto different phases of the theta oscillation,
this places a lower limit on the order of 100~ms on the timelines
indexed by $\taustar$ and $\delta$.  Theoretical and neurobiological considerations
led \citet{ShanEtal16} to the conclusion that $\delta$ ought to accelerate
exponentially with the theta cycle, resulting in a logarithmic sampling of the
$\delta$ axis.

This conjecture made sense of several neurophysiological findings, including
the gradual ramping of firing in striatal neurons accompanied by phase
precession with respect to theta recorded in the hippocampus, a brain
structure that is relatively distant from the striatum \citep{MeerEtal11}.
The fact that spikes in the striatum are organized by hippocampal theta
suggests that theta oscillations reflect a computation that is extended over a
significant part of the brain.  The learning rule presented here,
Eq.~\ref{eq:ppluspminus}, describes changes in the strength of connections in
$\C$ by noting the difference between $\pplus$ and $\pminus$ at each value of
$\delta$.     This suggests convergent connections between axons communicating
$\M$ and $\C$ arriving at target neurons representing predicted future
outcomes.    Perhaps the coordination implied by the involvement of theta
oscillations in prediction could lead to a difference in the timing of spikes
communicated \emph{via} $\M$ and $\C$.  Coupled with spike-timing-dependent
plasticity, perhaps this could lead to the learning rule in
Eq.~\ref{eq:ppluspminus}.

\section*{Acknowledgments}
This work was supported by NIBIB R01EB022864 and NSF IIS 1631460. The authors gratefully acknowledge inspiring conversations with Randy Gallistel and work in early stages of this project by Kostya Tiurev.

\subsection*{Code availability}
The code that supports the demonstration in Sec.~\ref{sec:Demonstration} can be found at \url{https://predicting.gitlab.io}.

\section*{Appendix}
\renewcommand{\thesection}{A}
\setcounter{subsection}{0}

\renewcommand\theequation{A\arabic{equation}}
\setcounter{equation}{0}

\subsection{A formal model for temporal record of the past \label{subsec:appendix-formal-model}}

Let multiple types of discrete events occur in continuous time. For each event type, we denote the signal by $f\left(t\right)$, where each event is represented by a Dirac delta function at the instant it occurs. For each event type, an array of leaky integrators, $F$, with a range of decay rates $s$, receive the signal as input:
\begin{equation}
\frac{\partial}{\partial t}F(s;t)=-sF(s;t)+f(t).
\label{eq:Fdiffeq}
\end{equation} 
The array of leaky integrators $F(s;t)$ encodes the real Laplace transform of the signal up to time $t$, where $s$ is the Laplace domain variable. For each event type, an array of time cells $\ftilde(\taustar)$ approximately inverts the Laplace transform \citep[see][]{Post30}. This yields an estimate of the signal up to time $t$, at time offsets $\taustar$ prior:
\begin{equation}
\ftilde(\taustar; t)=\ftilde(k/s;t)=\frac{(-1)^k}{k!}s^{k+1}  \frac{\partial^k}{\partial s^k}F(s;t)=\Lk F(s;t).
\label{eq:ftildedefining}
\end{equation} 
The constant $k$ is a sharpness parameter. As $k \rightarrow \infty$, the
estimate $\ftilde(\taustar)$ becomes precise, at the cost of infinite
resources to implement the model. As stated in Eq.~\ref{eq:ftildeintegral},
for an arbitrary signal $\mathbf{f}$,
\begin{equation}
	\ftildebold(\taustar; t) = \frac{1}{\taustar}\int_{-\infty}^t
	\mathbf{f}(\tau)\Phi_k\left(\frac{t-\tau}{\taustar}\right)\,d\tau .
    \label{eq:ftildeintegral'}
\end{equation}
In other words, for a given $\taustar$, $\ftildebold(\taustar; t)$ is
proportional to a causal convolution of the signal $\mathbf{f}$ with a kernel
$\Phi_k$ that describes the smearing.

\subsection{Time-translation to estimate the future state of the past}
\label{subsec:Appendix translation}

The future state of the memory (Eq.~\ref{eq:ftildedeltaintegral}) can be readily computed through translation in the Laplace domain:
\begin{equation}
	\ftildebold_\delta(\taustar;t) \equiv \Lk \mat{R}^\delta
	\mathbf{F}(s;t) \equiv \Lk \left\{e^{-s\delta} \mathbf{F}(s;t)\right\}.
    \label{eq:trans}
\end{equation}
Building a translation operator out of realistic neurons and synapses is a non-trivial,
but tractable problem.  It has been proposed that the brain implements
translation to various amounts $\delta$ by mapping $\delta$ on to different
phases of theta oscillations \citep{ShanEtal16}.  Previous work has long argued
that theta oscillations, a prominent 4--12~Hz oscillation in the local field potential,
have long been believed to be crucial in the neurobiology of memory
\citep{Buzs02,HassEtal02,KahaEtal01}.
Requiring scale invariance, and also consideration of the problem from the
perspective of the individual neurons requires the sweep through $\delta$ to
accelerate exponentially through the theta cycle.

\subsection{Pairwise association and pairwise prediction}
\label{subsec:Pairwise association and pairwise prediction}

The agent makes pairwise associations $\M$ between each pair
of event types using Hebbian learning. As the agent experiences the
world, the pairwise prediction $\mathbf{m}$ allows the agent to generate
predictions for the future based on pairwise associations with the
currently occurring events as cues. The pairwise prediction is a building
block for the learning of the credit associations $\exp\C$,
from which the prediction $\mathbf{p}$ is derived. This section consists
of two subsections. The first subsection motivates the form of the
pairwise prediction $\mathbf{m}$ (Eqs.~\ref{eq:m in terms of M operator}\textendash\ref{eq:fdelta};
in particular, the form of the integral in Eq.~\ref{eq:M operator definition}). The
second subsection highlights and proves the numerical coincidence
between $\M$ and $\mathbf{m}$ in a simple case, from which
the normalization for $\mathbf{m}$ derives.

\subsubsection{Equation for pairwise prediction}

In this subsection, we motivate the equations for the pairwise prediction.
We do this by showing that when memory is perfect, $\mathbf{m}$ reduces
to the geometric mean of the elements of $\M$ associated
with the events at time $t$, as desired.

Hebbian learning can be used to make pairwise associations $\M$
between events (Eq.~\ref{eq:M}). The parameters of $\M$ are
the possible cue, the possible outcome and the internal time. The
pairwise prediction $\mathbf{m}$ uses the pairwise associations $\mathbf{M}$
to make a prediction about future events based on possibly multiple
currently occurring events. In other words, $\mathbf{m}$ serves to
integrate cue\textendash outcome pairwise information from multiple
simultaneous cues in the present. $\M$, in the definition
for $\mathbf{m}$, plays the role that $\exp\C$ does in the
definition for $\mathbf{p}$. In the algorithm, $\mathbf{m}$ serves
the function of $\pplus$ (Eq.~\ref{eq:pplus}). 

The pairwise prediction $\mathbf{m}$ is computed, when a set of events
$\mathcal{E}_{t}$ occur at time $t$, as follows:

\begin{equation}
m^{\beta}\left(\delta;t\right)=\kappa_{1}\exp\left(\frac{1}{\left|\mathcal{E}_{t}\right|}\sum_{\alpha\in\mathcal{E}_{t}}\int f_{\delta}^{\alpha}(\overset{\ast}{\tau};t)\log M_{\alpha}^{\beta}(\overset{\ast}{\tau})\,d\overset{\ast}{\tau}\right),\qquad\delta>0,
\end{equation}
where the constant $\kappa_{1}=\left[ke^{-\psi\left(k\right)}\right]^{k+1}$,
$\psi\left(k\right)$ is the digamma function, and 
\[
f_{\delta}^{\alpha}(\overset{\ast}{\tau};t)=\Phi_{k}\left(\delta/\overset{\ast}{\tau}\right)/\overset{\ast}{\tau}
\]
denotes the future state of the memory element associated with the
currently occurring episode of $\alpha$. (For $k=2$, $\kappa_{1}\sim2.3$;
for $k=8$, $\kappa_{1}\sim1.8$; as $k\rightarrow\infty,$ $\kappa_{1}\rightarrow\sqrt{e}\sim1.65$.)
The notation $\left|\mathcal{E}_{t}\right|$ denotes the number of
elements in $\mathcal{E}_{t}$, that is, the number of events co-occurring
at time $t$. The function $f_{\delta}^{\alpha}(\overset{\ast}{\tau};t)$
is a Gaussian-like function that peaks around $\overset{\ast}{\tau}=\delta$,
and reflects the fact that at a time $\delta$ in the future, $\alpha$
would have occurred $\overset{\ast}{\tau}$ in the past, and the memory
$\tilde{f}^{\alpha}$ would reflect this.

We can motivate the form of the pairwise prediction $\mathbf{m}$
as follows. Let us imagine that memory were perfect such that events
were localized in time exactly, so $\phi_{\delta}^{\alpha}(\overset{\ast}{\tau})=\delta(\overset{\ast}{\tau}-\delta)I(\alpha\in\mathcal{E}_{t}),$
where $\phi_{\delta}^\alpha$ is the analog of $f_{\delta}^\alpha$ when memory
is perfect (i.e., $k\rightarrow\infty$), $\delta(\cdot)$ is the
Dirac delta function and $I(\cdot)$ is the indicator function. We
would then have 
\begin{align*}
\int\phi_{\delta}^{\alpha}(\overset{\ast}{\tau};t)\log M_{\alpha}^{\beta}(\overset{\ast}{\tau})\,d\overset{\ast}{\tau} & =I(\alpha\in\mathcal{E}_{t})\int\delta(\overset{\ast}{\tau}-\delta)\log M_{\alpha}^{\beta}(\overset{\ast}{\tau})\,d\overset{\ast}{\tau}\\
 & =I(\alpha\in\mathcal{E}_{t})\log M_{\alpha}^{\beta}(\delta).
\end{align*}
The integral on the right hand side is a convolution of $\log M_{\alpha}^{\beta}$
with the delta function, which returns the former unchanged. In the
case where exactly $\textsc{x}$ occurs at time $t$, 
\begin{align}
m^{\beta}\left(\delta;t\right) & \propto\exp\int f_{\delta}^{\alpha}(\overset{\ast}{\tau};t)\log M_{\alpha}^{\beta}(\overset{\ast}{\tau})\,d\overset{\ast}{\tau}\nonumber \\
 & =\exp\log M_{\mathrm{X}}^{\beta}(\delta)\nonumber \\
 & =M_{\mathrm{X}}^{\beta}(\delta),\label{eq:m=00003DM}
\end{align}
so $\mathbf{m}$ would be proportional to the appropriate elements
of $\M$. In the case where exactly $\textsc{x}$ and $\textsc{y}$
occur at time $t$, 
\begin{align*}
m^{\beta}\left(\delta;t\right) & \propto\exp\left[\frac{1}{\left|\mathcal{E}_{t}\right|}\sum_{\alpha\in\mathcal{E}_{t}}\int f_{\delta}^{\alpha}(\overset{\ast}{\tau};t)\log M_{\alpha}^{\beta}(\overset{\ast}{\tau})\,d\overset{\ast}{\tau}\right]\\
 & =\exp\left\{ \frac{1}{2}\left[\log M_{\mathrm{X}}^{\beta}(\delta)+\log M_{\mathrm{Y}}^{\beta}(\delta)\right]\right\} \\
 & =\sqrt{M_{\mathrm{X}}^{\beta}(\delta)M_{\mathrm{Y}}^{\beta}(\delta)},
\end{align*}
so $\mathbf{m}$ would be proportional to the geometric mean of the
elements of $\M$ associated with the events at time $t$. 

In the model with fuzzy memory, $f_{\delta}$ approximates $\phi_{\delta}$,
so the above relationships hold only approximately. In other words,
in general, according to Eqs.~\ref{eq:m in terms of M operator}\textendash\ref{eq:fdelta}, $m^{\beta}\left(\delta;t\right)$
is not always exactly $M_{\mathrm{X}}^{\beta}(\delta)$ (when $\textsc{x}$
occurs at time $t$) or $\sqrt{M_{\mathrm{X}}^{\beta}(\delta)M_{\mathrm{Y}}^{\beta}(\delta)}$
(when $\textsc{x}$ and $\textsc{y}$ occur at time $t$), and so
on. Instead of using Eqs.~\ref{eq:m in terms of M operator}\textendash\ref{eq:fdelta}, which involves
an integral, one could have defined $m^{\beta}\left(\delta;t\right)$
directly as the geometric mean of the relevant elements of $M^{\beta}(\delta)$.
However, we do not do so. We use Eqs.~\ref{eq:m in terms of M operator}\textendash\ref{eq:fdelta} because
they closely parallel Eqs.~\ref{eq:prediction}\textendash\ref{eq:C op definition} for the prediction,
for which the integral is necessary (see Sec.~\ref{subsec:Credit association and prediction}). Using
equations of similar form is more neurobiologically realistic, because
it suggests that analogous neural architecture supports the computation
for both pairwise prediction $\mathbf{m}$ and prediction $\mathbf{p}$.
Integrals in time are straightforward to implement with neural
networks.

In summary, we have shown that when memory is perfect, $\mathbf{m}$
reduces to the geometric mean of the elements of $\M$ associated
with the events at time $t$, as desired. The strength of the equations
underlying $\mathbf{m}$ is that they closely parallel those for $\mathbf{p}$. 

\subsubsection{Normalization for pairwise prediction}

In the bulk of the previous subsection, we imagined that memory were
perfect such that events were localized in time exactly. In the actual
formulation, memory is fuzzy, and this is reflected in the form of
$\textbf{f}_{\delta}$. Therefore, the above proportionality relationships
do not hold exactly, in general. However, Eq.~\ref{eq:m=00003DM}
holds at the time of occurrence of $\textsc{x}$ in the case where
$\textsc{x}$ and $\textsc{y}$ occur at a constant time interval
$\tau$, and no event co-occurs with $\textsc{x}$, even in the case
of fuzzy memory. In this subsection, we will prove this result (in
Lemma 3), which suggests the value that the normalization constant
$\kappa_{1}$ should take. Lemmas 1 and 2 are integrals that are used
to prove Lemma 3. 

\textbf{Lemma 1}. If the constant $a$ is positive and $k$ is a non-negative
integer,
\[
\int_{0}^{\infty}\frac{e^{-a/x}}{x^{k+2}}\,dx=\frac{k!}{a^{k+1}}.
\]

\textit{Proof}. The integral is 
\[
I=\int_{0}^{\infty}\frac{e^{-a/x}}{x^{k+2}}\,dx=\frac{1}{a}\int_{0}^{\infty}\frac{ae^{-a/x}}{x^{2}}x^{-k}\,dx.
\]
Noting that $\frac{d}{dx}\left(e^{-a/x}\right)=\frac{a}{x^{2}}e^{-a/x}$,
we integrate by parts: 
\begin{align*}
u & =x^{-k} & dv & =\frac{ae^{-a/x}}{x^{2}}\\
du & =\left(-k\right)x^{-k-1} & v & =e^{-a/x},
\end{align*}
so our integral is now
\[
I=\cancel{\left[\frac{1}{a}x^{-k}e^{-a/x}\right]_{0}^{\infty}}-\frac{-k}{a}\int_{0}^{\infty}\frac{e^{-a/x}}{x^{k+1}}\,dx=\frac{k}{a}\frac{1}{a}\int_{0}^{\infty}\frac{ae^{-a/x}}{x^{2}}x^{-\acute{\left(k-1\right)}}\,dx.
\]
If we were to integrate by parts again, we would have:
\begin{align*}
u & =x^{-\left(k-1\right)} & dv & =\frac{ae^{-a/x}}{x^{2}}\\
du & =-\left(k-1\right)x^{-k-1} & v & =e^{-a/x}.
\end{align*}
Each $i$th iteration reduces the exponent on $x$ in the denominator
of the integrand by 1 and introduces a factor of $\left(k-i+1\right)/a$,
and $k$ iterations are needed to go from having $x^{k+2}$ to $x^{2}$
in the denominator of the integrand. Noting that $\left(k\right)\left(k-1\right)\cdots\left(2\right)\left(1\right)=k!,$
we thus have
\[
I=\frac{k!}{a^{k}}\frac{1}{a}\int_{0}^{\infty}\frac{ae^{-a/x}}{x^{2}}\,dx=\frac{k!}{a^{k}}\frac{1}{a}\left[e^{-a/x}\right]_{0}^{\infty}=\frac{k!}{a^{k+1}}.\tag*{\ensuremath{\Box}}
\]

\textbf{Lemma 2}. If the constants $A$, $a$ and $b$ are positive
and $k$ and $m$ are positive integers, then
\begin{align*}
\int_{0}^{\infty}\frac{e^{-a/x}}{x^{k+1}}\log\frac{Ae^{-b/x}}{x^{m}}\,dx & =\frac{\left(k-1\right)!}{a^{k}}\left[m\psi\left(k\right)-\frac{kb}{a}+\log\frac{A}{a^{m}}\right],
\end{align*}
where $\psi\left(k\right)$ is the digamma function.

\textit{Proof}. The integrand is 
\[
\frac{e^{-a/x}}{x^{k+1}}\log\frac{Ae^{-b/x}}{x^{m}}=\frac{e^{-a/x}}{x^{k+1}}\log A-\frac{be^{-a/x}}{x^{k+2}}-m\frac{e^{-a/x}}{x^{k+1}}\log x.
\]
We integrate term by term. Applying Lemma 1, the first term is
\[
\log A\int_{0}^{\infty}\frac{e^{-a/x}}{x^{k+1}}\,dx=\left(\log A\right)\frac{\left(k-1\right)!}{a^{k}},
\]
noting that $k$ above is at least 1, and the second term is 
\[
-b\int_{0}^{\infty}\frac{e^{-a/x}}{x^{k+2}}\,dx=-b\frac{k!}{a^{k+1}}.
\]
The third term is 
\[
-m\int_{0}^{\infty}\frac{e^{-a/x}}{x^{k+1}}\log x\,dx,\qquad k\ge1.
\]
Substituting $t=a/x,$ $dt=-a/x^{2}\,dx$, we have
\begin{align*}
-\frac{m}{a^{k-1}}\int_{\infty}^{0}z^{k-1}e^{-t}\log\frac{a}{t}\left(-\frac{1}{a}\right)dt & =\frac{m}{a^{k}}\int_{0}^{\infty}t^{k-1}e^{-t}\log\frac{t}{a}\,dt\\
 & =\frac{m}{a^{k}}\left[\left(-\log a\right)\int_{0}^{\infty}t^{k-1}e^{-t}dt+\int_{0}^{\infty}t^{k-1}e^{-t}\log t\,dt\right]\\
 & =\frac{m}{a^{k}}\left[\left(-\log a\right)\Gamma\left(k\right)+\Gamma^{\prime}\left(k\right)\right]\\
 & =\frac{m}{a^{k}}\Gamma\left(k\right)\left[\psi\left(k\right)-\log a\right]\\
 & =\frac{m}{a^{k}}\left(k-1\right)!\left[\psi\left(k\right)-\log a\right],
\end{align*}
where we have applied Eqs.~5.2.1, 5.9.19 and 5.2.2 from \citet{OlveEtal21},
and used the fact that $k$ is a positive integer. Putting everything
together, we have
\begin{align*}
\int_{0}^{\infty}\frac{e^{-a/x}}{x^{k+1}}\log\frac{Ae^{-b/x}}{x^{m}}\,dx & =\left(\log A\right)\frac{\left(k-1\right)!}{a^{k}}-b\frac{k!}{a^{k+1}}+\frac{m}{a^{k}}\left(k-1\right)!\left[\psi\left(k\right)-\log a\right]\\
 & =\frac{\left(k-1\right)!}{a^{k}}\left\{ \log A-\frac{kb}{a}+m\left[\psi\left(k\right)-\log a\right]\right\} \\
 & =\frac{\left(k-1\right)!}{a^{k}}\left[m\psi\left(k\right)-\frac{kb}{a}+\log\frac{A}{a^{m}}\right].\tag*{\ensuremath{\Box}}
\end{align*}

\textbf{Lemma 3}. Let Event $i$ cue Event $j$ with a fixed time
interval $t_{ij}$, and let no event co-occur with Event $i$. After
training, at the time Event $i$ occurs,
\[
m^{j}\left(\delta\right)=M_{i}^{j}\left(\delta\right),
\]
for all $\delta>0.$

\textit{Proof}. On the right hand side, we have
\[
M_{i}^{j}\left(\delta\right)=K\frac{t_{ij}^{k}}{\delta^{k+1}}e^{-kt_{ij}/\delta},\qquad\delta>0,
\]
where $K=\frac{k^{k+1}}{k!}$ .

To compute the left hand side, we note that 
\[
f_{\delta}^{i}\left(\overset{\ast}{\tau}\right)=K\frac{\delta^{k}}{\overset{\ast}{\tau}^{k+1}}e^{-k\delta/\overset{\ast}{\tau}},\qquad\overset{\ast}{\tau},\delta>0.
\]
Then 
\begin{align}
\int f_{\delta}^{i}\left(\overset{\ast}{\tau}\right)\log M_{i}^{j}\left(\overset{\ast}{\tau}\right)\,d\overset{\ast}{\tau} & =\int_{0}^{\infty}K\frac{t_{ij}^{k}}{\overset{\ast}{\tau}^{k+1}}e^{-kt_{ij}/\overset{\ast}{\tau}}\log\left(K\frac{t_{ij}^{k}}{\overset{\ast}{\tau}^{k+1}}e^{-kt_{ij}/\overset{\ast}{\tau}}\right)\,d\overset{\ast}{\tau}\label{eq:where the proof breaks}\\
 & =K\delta^{k}\int_{0}^{\infty}\frac{e^{-k\delta/\overset{\ast}{\tau}}}{\overset{\ast}{\tau}^{k+1}}\log\left(\frac{Kt_{ij}^{k}}{\overset{\ast}{\tau}^{k+1}}e^{-kt_{ij}/\overset{\ast}{\tau}}\right)\,d\overset{\ast}{\tau}.
\end{align}
Substituting $a=k\delta$, $b=kt_{ij}$, $A=Kt_{ij}^{k}$, $m=k+1$
into Lemma 2, the above evaluates to
\begin{align*}
\int f_{\delta}^{i}\left(\overset{\ast}{\tau}\right)\log M_{i}^{j}\left(\overset{\ast}{\tau}\right)\,d\overset{\ast}{\tau} & =\cancel{K\delta^{k}\frac{\left(k-1\right)!}{\left(k\delta\right)^{k}}}\left[\left(k+1\right)\psi\left(k\right)-\frac{k^{2}t_{ij}}{k\delta}+\log\frac{Kt_{ij}^{k}}{\left(k\delta\right)^{k+1}}\right]\\
 & =\left(k+1\right)\psi\left(k\right)-\frac{kt_{ij}}{\delta}+\log\frac{Kt_{ij}^{k}}{\delta^{k+1}}-\left(k+1\right)\log k\\
 & =\log\frac{Kt_{ij}^{k}}{\delta^{k+1}}-\frac{kt_{ij}}{\delta}-\log\kappa_{1}
\end{align*}
Thus, the left hand side of the lemma is 
\begin{align*}
m^{j}\left(\delta\right) & =\kappa_{1}\exp\left(\frac{1}{\left|\mathcal{E}_{t}\right|}\sum_{\alpha\in\mathcal{E}_{t}}\int f_{\delta}^{\alpha}(\overset{\ast}{\tau};t)\log M_{\alpha}^{j}(\overset{\ast}{\tau})\,d\overset{\ast}{\tau}\right)\\
 & =\kappa_{1}\exp\left(\int f_{\delta}^{i}(\overset{\ast}{\tau};t)\log M_{i}^{j}(\overset{\ast}{\tau})\,d\overset{\ast}{\tau}\right)\\
 & =\kappa_{1}\exp\left(\log\frac{Kt_{ij}^{k}}{\delta^{k+1}}-\frac{kt_{ij}}{\delta}-\log\kappa_{1}\right)\\
 & =K\frac{t_{ij}^{k}}{\delta^{k+1}}e^{-kt_{ij}/\delta}\\
 & =M_{i}^{j}\left(\delta\right).\tag*{\ensuremath{\Box}}
\end{align*}

Our choice of $\kappa_{1}=\left[ke^{-\psi\left(k\right)}\right]^{k+1}$
allowed the equality $m^{j}\left(\delta\right)=M_{i}^{j}\left(\delta\right)$
to hold when Event $i$ cues Event $j$ with a fixed time interval
and no event co-occurs with Event $i$, without any additional constant
of proportionality in the equation.

\textit{Remark}. The conclusion of Lemma 3 does not hold in general
if the delay interval is not fixed. Let $i$ and then $j$ occur at
times 1 or 2 apart (with equal probability). Let $k=2$ (for ease
of calculation) so $K=\frac{k^{k+1}}{k!}=4$. We have
\begin{align*}
M_{i}^{j}\left(\overset{\ast}{\tau}\right) & =\frac{\int\tilde{f}_{\delta=0}^{i}\left(\overset{\ast}{\tau};t\right)f_{j}\left(t\right)\,dt}{\int f_{i}\left(t\right)\,dt}\\
 & =\frac{1}{2}\left(K\frac{1}{\overset{\ast}{\tau}^{k+1}}\left(e^{-k/\overset{\ast}{\tau}}+2^{k}e^{-2k/\overset{\ast}{\tau}}\right)\right)\\
 & =\frac{2}{\overset{\ast}{\tau}^{3}}\left(e^{-2/\overset{\ast}{\tau}}+4e^{-4/\overset{\ast}{\tau}}\right)\\
M_{i}^{j}\left(\overset{\ast}{\tau}=1\right) & =2\left(e^{-2}+4e^{-4}\right)=\boxed{0.417195},
\end{align*}
and as before 
\[
f_{\delta}^{i}\left(\overset{\ast}{\tau}\right)=4\frac{\delta^{2}}{\overset{\ast}{\tau}^{3}}e^{-2\delta/\overset{\ast}{\tau}}.
\]
But when $\delta=1$,
\begin{align*}
\int f_{\delta}^{i}\left(\overset{\ast}{\tau}\right)\log M_{i}^{j}\left(\overset{\ast}{\tau}\right)\,d\overset{\ast}{\tau} & =4\delta^{2}\int_{0}^{\infty}\frac{e^{-2\delta/\overset{\ast}{\tau}}}{\overset{\ast}{\tau}^{3}}\log\left[\frac{2}{\overset{\ast}{\tau}^{3}}\left(e^{-2/\overset{\ast}{\tau}}+4e^{-4/\overset{\ast}{\tau}}\right)\right]\,d\overset{\ast}{\tau}\\
 & =-1.51366\\
m^{j}\left(\delta=1\right) & =\kappa_{1}\exp\left(-1.51366\right)\\
 & =8\exp\left[-3\left(1-\gamma\right)\right]\exp\left(-1.51366\right)\\
 & =\boxed{0.495310}.
\end{align*}
Since the two boxed values are different, the equality $m^{j}\left(\delta\right)=M_{i}^{j}\left(\delta\right)$
does not hold in general.

\subsection{Credit association and prediction}
\label{subsec:Credit association and prediction}

The agent maintains credit associations $\exp\mathbf{C}$ between
each pair of event types, which estimates the multiplier for the agent's
belief about the rate of each outcome whenever it sees a potential
cue. The agent maintains predictions $\mathbf{p}$, a timeline of
future events based on the credit associations. As the agent experiences
the world, $\exp\mathbf{C}$ and $\mathbf{p}$ are iteratively updated. 

This section motivates the form of the prediction $\mathbf{p}$ (Eqs.~\ref{eq:prediction}--\ref{eq:C op definition}), by showing that the integral approximately
subtracts the time elapsed since each cue, from the time delay between
that cue and the outcome of interest. The integral thus produces a
function of $\delta$ that peaks approximately at the time remaining
to the outcome. This section consists of three subsections. In the
first subsection, we calculate the projected memory $\ftildebold_{\delta}$,
which is an element in the prediction $\mathbf{p}$, in the case of
perfect memory. In the second subsection, we calculate the prediction
$\mathbf{p}$ in the case of perfect memory. In the last subsection,
we discuss the case of fuzzy memory. 

\subsubsection{Perfect memory: Projected memory}

We proposed the following form for the prediction: 
\[
p^{\beta}\left(\delta;t\right)=\Lambda^{\beta}\exp\sum_{E\in\mathcal{E}}\int C_{E}^{\beta}(\overset{\ast}{\tau})\tilde{f}_{\delta}^{E}(\overset{\ast}{\tau};t)\,d\overset{\ast}{\tau},
\]
where $\Lambda^{\beta}$ denotes the long-term average of event type
$\beta$, and $\mathcal{E}$ denotes the set of possible cue types.
We motivate the above functional form by deriving the prediction in
the case of perfect memory (i.e., $k\rightarrow\infty$) and discrete
events. To do so, we would need to find the projected memory $\tilde{f}_{\delta}^{E}$
for event type $E$. Note (see Eq.~\ref{eq:impresp}) that 
\[
\lim_{k\rightarrow\infty}\Phi_{k}\left(x\right)=\delta\left(x-1\right),
\]
where $\delta\left(\cdot\right)$ represents the Dirac delta function.
Let the input function $\mathbf{f}$ be a series of discrete events
of type $e_{i}$ occurring at times $t_{i}<t$, 
\[
f^{E}\left(\tau\right)=\sum_{i}\delta\left(t_{i}-\tau\right)I\left(e_{i}=E\right).
\]
If we imagine that memory were perfect, then from Eq.~\ref{eq:ftildedeltaintegral},
the projected memory would be represented as 
\begin{align*}
\tilde{\phi}_{\delta}^{E}(\overset{\ast}{\tau};t) & =\lim_{k\rightarrow\infty}\frac{1}{\overset{\ast}{\tau}}\int_{-\infty}^{t}f^{E}\left(\tau\right)\Phi_{k}\left(\frac{t+\delta-\tau}{\overset{\ast}{\tau}}\right)\,d\tau\\
 & =\frac{1}{\overset{\ast}{\tau}}\int_{-\infty}^{t}\sum_{i}\delta\left(t_{i}-\tau\right)I\left(e_{i}=E\right)\delta\left(\frac{t+\delta-\tau}{\overset{\ast}{\tau}}-1\right)\,d\tau\\
 & =\sum_{i}I\left(e_{i}=E\right)\int_{-\infty}^{t}\delta\left(t_{i}-\tau\right)\frac{1}{\overset{\ast}{\tau}}\delta\left(\frac{t+\delta-\tau-\overset{\ast}{\tau}}{\overset{\ast}{\tau}}\right)\,d\tau\\
 & =\sum_{i}I\left(e_{i}=E\right)\int_{-\infty}^{t}\delta\left(t_{i}-\tau\right)\delta\left(t+\delta-\tau-\overset{\ast}{\tau}\right)\,d\tau\\
 & =\sum_{i}\delta(t-t_{i}+\delta-\overset{\ast}{\tau})I(e_{i}=E),
\end{align*}
where $\tilde{\phi}_{\delta}$ is the analog of $\ftildebold_{\delta}$
when memory is perfect, and we have used the property $\delta\left(\alpha x\right)=\delta\left(x\right)/\left|\alpha\right|$
of the Dirac delta function. 

\subsubsection{Perfect memory: Prediction}

We are now ready to derive the prediction:
\begin{align*}
p^{\beta}\left(\delta;t\right) & =\Lambda^{\beta}\exp\sum_{E\in\mathcal{E}}\int C_{E}^{\beta}(\overset{\ast}{\tau})\tilde{\phi}_{\delta}^{E}(\overset{\ast}{\tau};t)\,d\overset{\ast}{\tau}\\
 & =\Lambda^{\beta}\exp\sum_{i}\sum_{E\in\mathcal{E}}\int C_{E}^{\beta}(\overset{\ast}{\tau})\delta(t-t_{i}+\delta-\overset{\ast}{\tau})I(e_{i}=E)\,d\overset{\ast}{\tau}.
\end{align*}
Consider the summand associated with $i=1$,
\[
\sum_{E\in\mathcal{E}}\int C_{E}^{\beta}(\overset{\ast}{\tau})\delta(t-t_{1}+\delta-\overset{\ast}{\tau})I(e_{1}=E)\,d\overset{\ast}{\tau}=\int C_{e_{1}}^{\beta}(\overset{\ast}{\tau})\delta(t-t_{1}+\delta-\overset{\ast}{\tau})\,d\overset{\ast}{\tau}.
\]
The above integral is a convolution of $C_{e_{1}}^{\beta}(\delta)$
with $\delta_{-\left(t-t_{1}\right)}\left(\delta\right)=\delta(t-t_{1}+\delta)$.
The result is a translation of $C_{e_{1}}^{\beta}(\delta)$ by $-(t-t_{1})$,
the time interval since Event $i=1$:
\[
\int C_{e_{1}}^{\beta}(\overset{\ast}{\tau})\delta(t-t_{1}+\delta-\overset{\ast}{\tau})\,d\overset{\ast}{\tau}=C_{e_{1}}^{\beta}(t-t_{1}+\delta).
\]
This makes sense because $C_{e_{1}}^{\beta}$ peaks at the time that
$\beta$ is expected after having observed $e_{1}$ (say, $\tau$).
Since $e_{1}$ occurred $t-t_{1}$ ago, the view of $C_{e_{1}}^{\beta}$
that matters for the prediction should be translated by $-(t-t_{1})$.
This view peaks at $\tau-(t-t_{1})$, which is the time remaining
until $\beta$ is expected on the basis of $e_{1}.$ (For example,
if $\beta$ is expected 5 time units after $e_{1}$ ($C_{e_{1}}^{\beta}(\delta)$
peaks at $\delta=5$) and $e_{1}$ occurred 3 time units ago, then
$\beta$ is expected in 2 time units ($C_{e_{1}}^{\beta}(t-t_{1}+\delta)=C_{e_{1}}^{\beta}(3+\delta)$
peaks at $\delta=2$).) 

If there is only one event episode, indexed by $i=1$, 
\[
p^{\beta}\left(\delta;t\right)=\Lambda^{\beta}\exp C_{e_{1}}^{\beta}(t-t_{1}+\delta).
\]
Thus, the prediction $p^{\beta}$ will also peak at $\delta=\tau-(t-t_{1})$.
(Continuing with the example, the prediction will also peak at $\delta=2$.)
The form of the prediction makes sense. The quantity $\Lambda^{\beta}$
is the rate of occurrence of $\beta$ without any evidence, or the
agent's prior belief. The quantity $\exp C_{e_{1}}^{\beta}$ is the
adjustment factor for the rate of $\beta$ due to $e_{1}$. The product
is the new, adjusted estimate for the rate of $\beta$. 

If there are exactly two event types with one episode each, 
\[
p^{\beta}\left(\delta;t\right)=\Lambda^{\beta}\left[\exp C_{e_{1}}^{\beta}(t-t_{1}+\delta)\right]\left[\exp C_{e_{2}}^{\beta}(t-t_{2}+\delta)\right],
\]
and so on. In other words, the adjustment factors for different event
types multiply, as would be desired. 

\subsubsection{Fuzzy memory}

In the earlier subsections, we have assumed that memory is perfect
to illustrate the principles behind the form of the prediction $\mathbf{p}$.
However, memory is fuzzy in real, resource-bounded systems, so the
foregoing comments only apply approximately. The form of the integral
in the prediction, 
\[
\int C_{E}^{\beta}(\overset{\ast}{\tau})\tilde{f}_{\delta}^{E}(\overset{\ast}{\tau};t)\,d\overset{\ast}{\tau},
\]
is not a convolution since the width of $\tilde{f}_{\delta}^{E}$
increases with $\delta$. However, the purpose of the integral is
to approximate a convolution, so that the relevant view of $C_{E}^{\beta}$
can be used to generate a prediction. For example, if $E$ occurred
3 time units ago, $\tilde{f}_{\delta}^{E}(\overset{\ast}{\tau})$
would peak at $\overset{\ast}{\tau}=\delta+3$ units (as $\delta$
increases, $\tilde{f}_{\delta}^{E}(\overset{\ast}{\tau})$ becomes
increasingly wider). If $E$ cues $\beta$ with a delay of 5 time
units, $C_{E}^{\beta}(\overset{\ast}{\tau})$ would peak at around
5 time units. The integral attains its largest value approximately
when the peaks of $\tilde{f}_{\delta}^{E}(\overset{\ast}{\tau})$
and $C_{E}^{\beta}(\overset{\ast}{\tau})$ are aligned, which is around
$\delta=2.$ This makes sense, because $\beta$ is expected in 2 time
units from the present.

In summary, we have shown that the integral in Eq.~\ref{eq:C op definition} approximately
subtracts the time elapsed since each cue, from the time delay between
that cue and the outcome of interest. The integral thus produces a
function of $\delta$ that peaks approximately at the time remaining
to the outcome.

\subsection{Worked examples}
\label{subsec:A.5 Worked examples}

In Sec.~\ref{subsec:Equivalence of fuzzy memory and input temporal uncertainty}, we noted that fuzzy memory induces fuzziness
in the prediction $\mathbf{p}$, and noted that the latter could have
equivalently been induced by intrinsic temporal uncertainty in the
input in an agent with perfect memory, for every given snapshot in
time. In Sec.~\ref{subsec:With fuzzy memory, credit is assigned based on temporal proximity}, we noted that other things equal, cues
closer in time to outcomes tend to receive more credit for those outcomes,
in the case of fuzzy memory. We provided one example each for illustration.
This section consists of two subsections. In the first subsection,
we provide the mathematical details for the example in Sec.~\ref{subsec:Equivalence of fuzzy memory and input temporal uncertainty}.
In the second subsection, we do so for the example in Sec.~\ref{subsec:With fuzzy memory, credit is assigned based on temporal proximity}.

\subsubsection{Worked example 1: Forward conditioning}
\label{subsec:A.5.1 Worked example 1: Forward conditioning}

For this example, Event $\textsc{x}$ cues Event $\textsc{y}$ with
a fixed delay of time interval $\tau$. We work out the agent's prediction
for $\textsc{y}$ at the time $\textsc{x}$ occurs. 

The pairwise association is given in Eq.~\ref{eq:M in XY case},
\[
M_{\mathrm{X}}^{\mathrm{Y}}(\overset{\ast}{\tau})=\Phi_{k}(\tau/\overset{\ast}{\tau})/\overset{\ast}{\tau}.
\]
By Lemma 3, when $\textsc{x}$ occurs, 
\[
m^{\mathrm{Y}}(\overset{\ast}{\tau})=M_{\mathrm{X}}^{\mathrm{Y}}(\overset{\ast}{\tau})=\Phi_{k}(\tau/\overset{\ast}{\tau})/\overset{\ast}{\tau}.
\]
When $\textsc{x}$ occurs, the memory is empty, so 
\[
	(p^{-})^\mathrm{Y}=\Lambda^{\mathrm{Y}}.
\]
Thus, after training, the credit due to $\textsc{x}$ for $\textsc{y}$
is
\[
\exp C_{\mathrm{X}}^{\mathrm{Y}}(\overset{\ast}{\tau})=\frac{p^{+}(\overset{\ast}{\tau})}{p^{-}(\overset{\ast}{\tau})}=\frac{\Phi_{k}(\tau/\overset{\ast}{\tau})}{\Lambda^{\mathrm{Y}}\overset{\ast}{\tau}}.
\]
The projected memory for $\textsc{x}$ when $\textsc{x}$ occurs is
\[
\tilde{f}_{\delta}^{\mathrm{X}}(\overset{\ast}{\tau})=\Phi_{k}(\delta/\overset{\ast}{\tau})/\overset{\ast}{\tau}.
\]
The prediction for $\textsc{y}$ at the time $\textsc{x}$ occurs
is 
\begin{align}
p^{\mathrm{Y}}(\delta) & =\Lambda^{\mathrm{Y}}\exp\sum_{E\in\mathcal{E}}\int C_{E}^{\mathrm{Y}}(\overset{\ast}{\tau})\tilde{f}_{\delta}^{E}(\overset{\ast}{\tau})\,d\overset{\ast}{\tau}\nonumber \\
 & =\Lambda^{\mathrm{Y}}\exp\int C_{\mathrm{X}}^{\mathrm{Y}}(\overset{\ast}{\tau})\tilde{f}_{\delta}^{\mathrm{X}}(\overset{\ast}{\tau})\,d\overset{\ast}{\tau}.\label{eq:pY worked example}
\end{align}
The integral is 
\begin{align}
\int C_{\mathrm{X}}^{\mathrm{Y}}(\overset{\ast}{\tau})\tilde{f}_{\delta}^{\mathrm{X}}(\overset{\ast}{\tau})\,d\overset{\ast}{\tau} & =\int\frac{\Phi_{k}(\delta/\overset{\ast}{\tau})}{\overset{\ast}{\tau}}\log\frac{\Phi_{k}(\tau/\overset{\ast}{\tau})}{\Lambda^{\mathrm{Y}}\overset{\ast}{\tau}}\,d\overset{\ast}{\tau}\label{eq:Cf integral worked example}\\
 & =\kappa_{0}\delta^{k}\int\frac{e^{-k\delta/\overset{\ast}{\tau}}}{\overset{\ast}{\tau}^{k+1}}\log\frac{\kappa_{0}\tau^{k}e^{-k\tau/\overset{\ast}{\tau}}}{\Lambda^{\mathrm{Y}}\overset{\ast}{\tau}^{k+1}}\,d\overset{\ast}{\tau}.\nonumber 
\end{align}
By Lemma 2, we have 
\begin{align*}
\int_{0}^{\infty}\frac{e^{-a/x}}{x^{k+1}}\log\frac{Ae^{-b/x}}{x^{m}}\,dx & =\frac{\left(k-1\right)!}{a^{k}}\left[m\psi\left(k\right)-\frac{kb}{a}+\log\frac{A}{a^{m}}\right],
\end{align*}
so we substitute $a=k\delta$, $A=\kappa_{0}\tau^{k}/\Lambda^{\mathrm{Y}}$,
$b=k\tau$ and $m=k+1$ to find
\[
\int_{0}^{\infty}\frac{e^{-k\delta/\overset{\ast}{\tau}}}{\overset{\ast}{\tau}^{k+1}}\log\frac{\kappa_{0}\tau^{k}e^{-k\tau/\overset{\ast}{\tau}}}{\Lambda^{\mathrm{Y}}\overset{\ast}{\tau}^{k+1}}\,d\overset{\ast}{\tau}=\frac{\left(k-1\right)!}{\left(k\delta\right)^{k}}\left[\left(k+1\right)\psi\left(k\right)-\frac{k\tau}{\delta}+\log\frac{\kappa_{0}\tau^{k}}{\Lambda^{\mathrm{Y}}\left(k\delta\right)^{k+1}}\right].
\]
Therefore, 
\begin{align*}
\int C_{\mathrm{X}}^{\mathrm{Y}}(\overset{\ast}{\tau})\tilde{f}_{\delta}^{\mathrm{X}}(\overset{\ast}{\tau})\,d\overset{\ast}{\tau} & =\kappa_{0}\delta^{k}\frac{\left(k-1\right)!}{\left(k\delta\right)^{k}}\left[\left(k+1\right)\psi\left(k\right)-\frac{k\tau}{\delta}+\log\frac{\tau^{k}}{\Lambda^{\mathrm{Y}}k!\delta^{k+1}}\right]\\
 & =\left(k+1\right)\psi\left(k\right)-\frac{k\tau}{\delta}+\log\frac{\tau^{k}}{\Lambda^{\mathrm{Y}}k!\delta^{k+1}}
\end{align*}
and the prediction for $\textsc{y}$ at the time $\textsc{x}$ occurs
is 
\begin{align*}
p^{\mathrm{Y}}(\delta) & =\Lambda^{\mathrm{Y}}\exp\int C_{\mathrm{X}}^{\mathrm{Y}}(\overset{\ast}{\tau})\tilde{f}_{\delta}^{\mathrm{X}}(\overset{\ast}{\tau})\,d\overset{\ast}{\tau}\\
 & =\Lambda^{\mathrm{Y}}\exp\left[\left(k+1\right)\psi\left(k\right)-\frac{k\tau}{\delta}+\log\frac{\tau^{k}}{\Lambda^{\mathrm{Y}}k!\delta^{k+1}}\right]\\
 & =\Lambda^{\mathrm{Y}}\exp\left[\left(k+1\right)\psi\left(k\right)\right]\frac{\tau^{k}}{\Lambda^{\mathrm{Y}}k!\delta^{k+1}}e^{-k\tau/\delta}\\
 & =\frac{\exp\left[\left(k+1\right)\psi\left(k\right)\right]}{k!}\frac{\tau^{k}}{\delta^{k+1}}e^{-k\tau/\delta}\\
 & =\frac{\exp\left[\left(k+1\right)\psi\left(k\right)\right]}{k^{k+1}}\frac{1}{\delta}\kappa_{0}\left(\frac{\tau}{\delta}\right)^{k}e^{-k\tau/\delta}\\
 & =\frac{\kappa_{1}^{-1}}{\delta}\Phi_{k}\left(\frac{\tau}{\delta}\right).
\end{align*}

This equation is Eq.~\ref{eq:fuzzymemoryprediction}. Sec.~\ref{subsec:Equivalence of fuzzy memory and input temporal uncertainty} makes use of
this result. 

\subsubsection{Worked example 2: Credit and temporal proximity}
\label{subsec:A.5.2 Worked example 2: Credit and temporal proximity}

For this example, Event $\textsc{x}$ cues Event $\textsc{y}$ and
Event $\textsc{z}$ with a fixed delay of time interval $2-t_{\mathrm{YZ}}$
and 2 respectively (relative to the time of occurrence of $\textsc{x}$).
We work out $\exp C_{\mathrm{Y}}^{\mathrm{Z}}$ as a function of $t_{\mathrm{YZ}}$.
The significance of this result is discussed in Sec.~\ref{subsec:With fuzzy memory, credit is assigned based on temporal proximity}. 

We need $\left(p^{+}\right)_{\mathrm{Y}}^{\mathrm{Z}}$, which we write as
shorthand for ``the degree to which $\textsc{z}$ appears in $\pplus$ due to $\textsc{y}$'' and $\left(p^{-}\right)^{\mathrm{Z}}$
at the time $\textsc{y}$ occurs. 
\[
\left(p^{-}\right)^{\mathrm{Z}}=\Lambda^{\mathrm{Z}}\exp\sum_{E\in\mathcal{E}}\int C_{E}^{\mathrm{Z}}(\overset{\ast}{\tau})\tilde{f}_{\delta}^{E}(\overset{\ast}{\tau};t_{\mathrm{Y}}^{-})\,d\overset{\ast}{\tau},
\]
where $t_{\mathrm{Y}}^{-}$ refers to the moment just before $\textsc{y}$
occurs. At that time, the memory contains only $\textsc{x}$ which
occurred $(2-t_{\mathrm{YZ}}$) ago, so
\begin{equation}
\left(p^{-}\right)^{\mathrm{Z}}=\Lambda^{\mathrm{Z}}\exp\int C_{\mathrm{X}}^{\mathrm{Z}}(\overset{\ast}{\tau})\tilde{f}_{\delta}^{\mathrm{X}}(\overset{\ast}{\tau};t_{\mathrm{Y}}^{-})\,d\overset{\ast}{\tau}.\label{eq:pminus Z worked example}
\end{equation}
The credit due to $\textsc{x}$ for $\textsc{z}$ is 
\[
\exp C_{\mathrm{X}}^{\mathrm{Z}}(\delta)=\frac{\Phi_{k}(2/\overset{\ast}{\tau})}{\Lambda^{\mathrm{Z}}\overset{\ast}{\tau}},
\]
and the projected memory for $\mathrm{\textsc{x}}$ when $\textsc{y}$
occurs is
\[
\tilde{f}_{\delta}^{\mathrm{X}}(\overset{\ast}{\tau})=\Phi_{k}\left((\delta+2-t_{\mathrm{YZ}})/\overset{\ast}{\tau}\right)/\overset{\ast}{\tau}.
\]
Thus, the integral is 
\begin{align*}
\int C_{\mathrm{X}}^{\mathrm{Z}}(\overset{\ast}{\tau})\tilde{f}_{\delta}^{\mathrm{X}}(\overset{\ast}{\tau};t_{\mathrm{Y}}^{-})\,d\overset{\ast}{\tau} & =\int\frac{1}{\overset{\ast}{\tau}}\Phi_{k}\left(\frac{\delta+2-t_{\mathrm{YZ}}}{\overset{\ast}{\tau}}\right)\log\frac{\Phi_{k}(2/\overset{\ast}{\tau})}{\Lambda^{\mathrm{Z}}\overset{\ast}{\tau}}\,d\overset{\ast}{\tau}.
\end{align*}
This integral is the same as Eq.~\ref{eq:Cf integral worked example}
in the previous example with $\delta$, $\tau$ and $\Lambda^{\mathrm{Y}}$
replaced with $\delta+2-t_{\mathrm{YZ}}$, 2 and $\Lambda^{\mathrm{Z}}$
respectively. The same relationship holds between $\left(p^{-}\right)^{\mathrm{Z}}$
here (Eq.~\ref{eq:pminus Z worked example}) and $p^{\mathrm{Y}}$
in the previous example (Eq.~\ref{eq:pY worked example}). Referring
to the previous result, we thus have
\[
\left(p^{-}\right)^{\mathrm{Z}}(\delta)=\frac{\kappa_{1}^{-1}}{\delta+2-t_{\mathrm{YZ}}}\Phi_{k}\left(\frac{2}{\delta+2-t_{\mathrm{YZ}}}\right).
\]
On the other hand, 
\[
\left(p^{+}\right)_{\mathrm{Y}}^{\mathrm{Z}}(\delta)=m^{\mathrm{Z}}(\delta)=M_{\mathrm{Y}}^{\mathrm{Z}}(\delta)=\Phi_{k}(t_{\mathrm{YZ}}/\delta)/\delta.
\]
Thus, after training, the credit due to $\textsc{y}$ for $\textsc{z}$
is 
\begin{align*}
\exp C_{\mathrm{Y}}^{\mathrm{Z}}(\delta) & =\frac{\left(p^{+}\right)_{\mathrm{Y}}^{\mathrm{Z}}(\delta)}{\left(p^{-}\right)^{\mathrm{Z}}(\delta)}\\
 & =\frac{\Phi_{k}(t_{\mathrm{YZ}}/\delta)/\delta}{\frac{\kappa_{1}^{-1}}{\delta+2-t_{\mathrm{YZ}}}\Phi_{k}\left(\frac{2}{\delta+2-t_{\mathrm{YZ}}}\right)}\\
 & =\frac{\left(t_{\mathrm{YZ}}/\delta\right)^{k}\exp\left(-kt_{\mathrm{YZ}}/\delta\right)/\delta}{\frac{\kappa_{1}^{-1}}{\delta+2-t_{\mathrm{YZ}}}\left(\frac{2}{\delta+2-t_{\mathrm{YZ}}}\right)^{k}\exp\left(-\frac{2k}{\delta+2-t_{\mathrm{YZ}}}\right)}\\
 & =\kappa_{1}\left(\frac{t_{\mathrm{YZ}}}{2}\right)^{k}\left(\frac{\delta+2-t_{\mathrm{YZ}}}{\delta}\right)^{k+1}\exp\left(-k\left[\frac{t_{\mathrm{YZ}}}{\delta}-\frac{2}{\delta+2-t_{\mathrm{YZ}}}\right]\right).
\end{align*}
Fig.~\ref{fig:f5} plots this expression for various values of $t_{\mathrm{YZ}}$.
See Sec.~\ref{subsec:With fuzzy memory, credit is assigned based on temporal proximity} for a discussion.

\subsection{Demonstration: Methods}
\label{subsec:Methods for demonstration}

Given a time-ordered set of events $[e_1,e_2,...,e_n]$, where each $e_i =
(x_i,t_i)$ comprises a discrete-valued type and real-valued timestamp, 
we are interested in predicting the type $x_{n+1}$ of the next event given its time of
occurrence $t_{n+1}$.
In the demonstration, we apply the prediction algorithm (``$\mathcal{C}$-based'') to a
{\it superposition} of independent MRPs and compare its predictions to those of a pairwise event association model (``$\mathcal{M}$-based'').
In the simulation, both the $\mathcal{C}$- and $\mathcal{M}$-based predictors have memories spanning $10^{-5}$ to $80$~time units into the past, each covered by 200 log-spaced memory nodes. 

Within each MRP, the probability of the type {\it and} time of an event depends solely on the type of the most recent past event, i.e., for MRP $k$,
\begin{equation*} {P}((x^k_{n+1}, t^k_{n+1})
    |\{(x^k_m,t^k_m)\}_{m\leq n}) = {P}((x^k_{n+1}, t^k_{n+1}) |x^k_n),
\end{equation*}
where $t^k_{n+1} > t^k_n$. The set of event types within each MRP is discrete and finite, while transition times $\Delta t_{n+1} = t_{n+1} - t_n > 0$ are real and strictly positive; this allows only one event to occur at a given time. 
Within each MRP, the probability of the type of the next event is given by the transition matrix 
\begin{equation*}
P_{ij} = P(x^k_{n+1}=j|{x^k_{n}=i})=
    \begin{pmatrix} 
0.05 & 0.75 & 0.2\\ 
0.2 & 0.05 & 0.75\\ 
0.75 & 0.2 & 0.05
\end{pmatrix}.
\end{equation*}
The transition times from $i$ to $j$ in MRP $k$ follows a truncated normal distribution $\mathcal{N}(\mu^k_{ij},\sigma^{2k}_{ij})$, with a lower bound cutoff of $10^{-5}$ (to ensure positivity).

We use two approaches which generate superposed processes differently.
We discuss the first approach, used for Fig.~\ref{fig:processesnwalkers}c.
The means $\mu^k_{ij}$ and variances $\sigma^{2k}_{ij}$ of the transition time distributions are drawn uniformly from the intervals $(0,10)$ and $(0,2)$, respectively. The same values are used across all six runs of the simulation. For each run, we generate exactly seven MRPs, labeled $k=1,\dots,7$, each with 500 event episodes. We then construct seven superposed processes from the aforementioned MRPs as follows. The first superposed process consists of one MRP, namely, the MRP $k=1$; the second superposed process consists of two MRPs, namely the MRPs with $k=1$ and $k=2$; and so on. Each component MRP has three types of events, so the total number of event types in the superposed process is $3N$, where $N$ is the number of MRPs superposed. 
 
We now discuss the second approach, used for Fig.~\ref{fig:processesnwalkers}d.
We draw exactly one set of transition time distribution parameters $\mu_{ij}$ and $\sigma^2_{ij}$ as before. This same set of parameters is used across all six runs of the simulation, and for all MRPs $k=1,\dots,7$. We generate exactly seven MRPs of 20,000 event episodes each, and generate seven superposed processes therefrom by incrementally superposing the MRPs as in the first approach. 
Every MRP has three types of events ($\e{6},\e{7},\e{8}$). 
In the superposed processes, the event types are not distinguished according to the MRP of origin (e.g., a $\e{6}$ from one MRP and a $\e{6}$ from another MRP are both of type $\e{6}$ in the superposed process).
Thus, in contrast to the previous approach, the algorithms only observe three types of events in the superposed MRPs.

In both Fig.~\ref{fig:processesnwalkers}c and \ref{fig:processesnwalkers}d, 80\% of each superposed process is used for training and the rest for testing.
For the $\mathcal{C}$-based prediction, accuracy on the test set is computed by checking if, at every time $t_n$ that an event occurs, the prediction evaluated at $t_{n+1}$, the time  of the next event, $\text{argmax}_i ~ p^i(\delta=\Delta t_{n+1};t=t_n)$ matches the event that actually occurs at that time. 
For the $\mathcal{M}$-based prediction, the computation is analogous, except the prediction is found \textit{via} $\text{argmax}_i ~ m^i(\delta=\Delta t_{n+1};t=t_n)$, where $j=x_n$, the type of the event at $t_n$.
The simulation is run 6 times and the average accuracy is reported.

\bibliographystyle{APA}
\bibliography{bibdesk,citations}

\end{document}